\documentclass[useAMS,usenatbib,usegraphicx,onecolumn,referee]{mn2e} % tasta saa one/two column, referee=one column
\title[NIRS0S atlas]{Near-IR Atlas of S0-Sa galaxies (NIRS0S)}

\author[E-mail:eija.laurikainen@oulu.fi]{E. Laurikainen$^{1,2}$\thanks{E-mail:eija.laurikainen@oulu.fi}, H. Salo$^{1}$, R. Buta$^{3}$, J. H. Knapen$^{4,5}$\\
$^{1}$Dept. of Physics/Astronomy Division, University of Oulu, FI-90014 Finland\\
$^{2}$Finnish Centre of Astronomy with ESO (FINCA), University of Turku, V\"ais\"al\"antie 20, FI-21500, Piikki\"o, Finland\\ 
$^{3}$Department of Physics and Astronomy, University of Alabama, Box 870324, Tuscaloosa, AL 35487\\
$^{4}$Instituto de Astrof\'\i sica de Canarias, E-38200 La Laguna, Tenerife, Spain\\
$^{5}$Departamento de Astrof\'isica, Universidad de La Laguna, E-38205 La Laguna, Tenerife, Spain}
\begin{document}

\date{Accepted: Received:}

%%%%\pagerange{\pageref{firstpage}--\pageref{lastpage}} \pubyear{2007}

\maketitle

\label{firstpage}
 
\begin{abstract}

An atlas of $K_{\rm s}$-band images of 206 early-type galaxies is
presented, including 160 S0-S0/a galaxies, 12 ellipticals, and 33 Sa
galaxies (+ one later type).  A majority of the Atlas galaxies
  belong to a magnitude-limited (m$_B$$\leq$12.5 mag) sample of 185
  NIRS0S (Near-IR S0 galaxy Survey) galaxies. To assure that
mis-classified S0s are not omitted, 25 ellipticals from RC3 classified
as S0s in the Carnegie Atlas were included in the sample.  The
observations were carried out using 3-4 m class telescopes with
sub-arcsecond pixel resolution ($\sim$ 0.25''), and were obtained in
good seeing conditions (FWHM $\sim$ 1'').  The images are 2-3 mag
deeper than 2MASS images, allowing the detection of faint outer disks
in S0s. Both {\it visual} and {\it photometric} classifications are
made, largely following the classification criteria of
\citet{vaucouleurs1959}.  Special attention is paid to the
classification of lenses, which are coded in a more systematic manner
than in any of the previous studies. A new lens-type, called a
'barlens', is introduced, possibly forming part of the bar
itself. Also, boxy/peanut/x-shaped structures 
are identified in many barred galaxies, even-though the galaxies are not seen
in edge-on view, indicating that vertical thickening is not enough 
to explain these structures, indicating that vertical thickening
is not enough to explain them. Photometric classification includes detection of
exponential outer disks or other structures not directly visible in
the images, but becoming clear in unsharp masking or residual images
in decompositions. In our photometric classification, nuclear bars are
assigned for 15 galaxies, which are overshadowed by bulges in visual
classification.  The mean Hubble stage is found to be similar in the
near-IR and in the optical.  We give dimensions of structure
components, and radial profiles of the position angles and
ellipticities, and show deviations from perfect elliptical isophotes.
Shells and ripples, generally assumed to be manifestations of recent
mergers, are detected only in 6 galaxies. However, multiple lenses
appear in as much as 25$\%$ of the Atlas galaxies, which is a
challenge to the hierarchical evolutionary picture of galaxies. Such
models need to explain how the lenses were formed, and then survived
in multiple merger events that galaxies may have suffered during their
lifetimes.

\end{abstract}


\section{INTRODUCTION}

In the early classification by \citet{hubble1936}, the S0s were an
enigmatic group of galaxies between the ellipticals and the early-type
spirals, and since then they have been subject to many kinds of
interpretations.  The classification of S0s depends on recognizing the
presence of a disk, but having no spiral arms. The interface between E
and S0 galaxies was somewhat obscured by the detection of boxy and
disky ellipticals \citep{bender1988}, based on deviations of the outer
isophotes from simple elliptical shape.  Kinematic observations by
Dressler $\&$ Sandage (1983) had already shown earlier that the lower
luminosity ellipticals are more rotationally supported than the bright
ellipticals. Bender et al. (1989) then discovered that both galaxy
luminosity and the degree of rotational support correlate with the
isophotal shapes of the elliptical galaxies.  A similar sequence of
increasing dominance of rotational support towards the
lower-luminosity galaxies was found also for the S0s
\citep{dressler1983}. However, as the amount of rotation is
significantly larger in S0s, this kinematically links the S0s to the
spiral galaxies. All this caused some early-type galaxy observers to
question the morphological classification of S0s, an attitude which 
culminated in 1990 when King (1992) and Djorgovski (1992) announced
that the Hubble sequence was breaking down, and should be replaced by
a classification based on measured physical parameters.  Despite the
early discovery of S0s, new kinematic observations have put them once
again at the forefront of research. Recent IFU-kinematic observations
by \citet{emsellem2007} have shown that the fast rotators are
morphologically assigned to a mix of E and S0 galaxies, which leads to
further questions about the meaning of their morphological
classification.

On the other hand, there are morphological structures in S0s which can
be connected successfully to real dynamical processes. Inner, outer
and nuclear rings can be either bar-induced resonance rings, or
accretion rings related to accumulation of external gas into the
galactic disks (see review by Buta 2011).  Recently new theories of
ring formation have also been presented, like the ``manifold orbits''
emanating from Lagrangian points in barred potentials
\citep{romero2006,atha2010}. Merger-built structures such as shells,
ripples \citep{malin1980} and polar rings (Schweizer et al. 1983; in
2$\%$ of S0s) appear in S0 galaxies, but they are not very common. The
first attempt to include bar morphology, e.g., boxy, peanut or
x-shaped structures into galaxy classification was made by
\citet{butaetal2010}. Bar morphology is an important characteristic of
galaxy morphology, which in theoretical models has been associated
with secular dynamical evolution of galaxies
\citep{atha2002,atha2003}.

Lenses formed part of the early classification scheme of S0s
\citep{sandage1961,sandage1981}, but were not initially assigned any
classification symbols. Inner and outer lenses in barred S0s were
discussed by \citet{kormendy1979}, and \citet{lauri2009} showed that a
large majority of S0s, both barred and non-barred, have lenses.
Moreover, some S0s have complicated multi-lens systems, which have not
yet been theoretically explained. Overall, the origin of lenses is not
well understood: they can form as part of the disk formation process
\citep{bosma1983}, be triggered by bars \citep{kormendy1979}, or by
the accretion of small companions. In fact, lenses and other
fine-structures of S0s might be important imprints of possible secular
evolution of galaxies. The only major galaxy atlas that recognizes
lenses in the classification is the de Vaucouleurs Atlas of Galaxies
(Buta et al. 2007; hereafter dVA).

In this paper we present the NIRS0S (Near-IR S0 galaxy Survey) atlas
in the $K_{\rm s}$-band, and use it for detailed morphological
classification. To our knowledge, this is the first attempt of
detailed classification of S0s using deep near-IR images.  The 2.2
$\mu m$ wavelength used traces the old stellar population of galaxies
and is relatively free of internal extinction, which makes it ideal
for the classification of structures. The sample was selected
  from the Third Reference Catalogue of Bright Galaxies (de
  Vaucouleurs et al. 1991; hereafter RC3).  In order to study the interfaces of
  S0s with ellipticals and spirals, Sa galaxies in RC3, and those ellipticals
  classified as S0s in the Revised Shapley-Ames Catalogue of Bright
  Galaxies (Sandage $\&$ Tammann 1981; RSA), were also included in the
  sample. Our images are several magnitudes deeper than the images in
the Two-Micron All-Sky Survey (2MASS, Skrutskie et al. 2006), which is
the largest near-IR survey obtained previously. Two large mid-IR
galaxy surveys using the {\it Spitzer Space Telescope} are the Spitzer
Infrared Nearby Galaxies Survey (SINGS; Kennicutt et al. 2003), and
the Local Volume Legacy project (LVL; Kennicutt et al. 2007), both
providing deep images at 3.6 $\mu m$. However, these surveys contain
only a few S0s.  A more comprehensive nearby galaxy survey is the {\it
  Spitzer} Survey of Stellar Structure in galaxies (S$^4$G; Sheth et
al. 2010), which consists of 2331 nearby galaxies. This survey exceeds
the image depth of NIRS0S, but NIRS0S is more complete in respect of
the S0s, and the pixel resolution is higher than in S$^4$G.

The NIRS0S atlas consists of images of 206 galaxies, a sample which,
after our revised classification, has 12 ellipticals, 160 S0-S0/a, and
33 Sa galaxies. Section 2 describes the sample and observations, data
reductions are explained in Section 3, and the image atlas in Section
4.  Visual and photometric classifications are presented, starting from
the de Vaucouleurs' (1959) classification criteria, but going beyond
that in classifying the detail of structures (Section 5). The
dimensions of the structure components are given in Section 6, and the
radial profiles of the position angles, ellipticities, and of the
parameter $b4$, describing deviations from perfect ellipticity of the
isophotal contours, are shown in the Atlas (Fig. 5).  In this paper,
the image Atlas is presented, whereas the number statistics and more
thorough discussion of the structure components will appear in 
forthcoming papers.

We find that multiple lenses are common in S0s, appearing even in
25$\%$ of the Atlas galaxies. However, shells or ripples were
detected only in 6 galaxies. Of the 25 RC3 ellipticals in our
original sample, 7 were re-classified as S0s by us.  Bars and bulges
in sub-samples of NIRS0S have been previously discussed by
\citet{lauri2005,lauri2006,lauri2007,lauri2009} and \citet{butaetal2006}, the
properties of bulges by \citet{lauri2010}, and the distribution of bar
strengths by \citet{butaetal2010}.

\section{SAMPLE AND OBSERVATIONS}

We have carried out a large, magnitude-limited imaging survey, the
Near-IR S0 galaxy Survey (NIRS0S) in the nearby Universe.  The sample
selection criteria are as follows: morphological type $-$3 $\leq$ $T$ $\leq$ 1, 
total magnitude of $B_{\rm T}$ $\le$ 12.5 mag, and inclination less than
65$^o$.  Applying these criteria to RC3, and including also
25 ellipticals (including late-type E$^+$) classified as S0s in RSA, yields a
sample of 185 galaxies (marked with an asterisk in Table 3).\footnote{Our
  current NIRS0S sample differs from that specified by
  \citet{lauri2005} and \citet{butaetal2006} in that we use $B_{\rm T}$ or the
  photographic value $m_{\rm B}$, or the average of these two when both are
  available. This was done to eliminate contamination of the original
  sample by total $V$ magnitudes in RC3, which occupy the same column
  as $B_{\rm T}$ in that catalogue.}  These ellipticals were included, in order
not to miss any potentially misclassified S0s. The sample includes 30 additional
galaxies not fulfilling the original selection criteria, mostly
S0-Sa galaxies which slightly exceed the magnitude limit, or in some
cases the inclination limit.  These galaxies were observed when it was
not possible to observe the primary targets, due to unsuitable wind
direction, or because no primary targets were visible. Including 
these galaxies yields a sample of 215 galaxies. In total, after our
re-classifications, the full sample includes 13 ellipticals, 139 S0s, 30
S0/a galaxies, 33 Sa galaxies and one later-type spiral.  The
selection criteria in our magnitude-limited NIRS0S sample are similar
to those in the Ohio State University Bright Spiral Galaxy Survey
(Eskridge et al. 2002; OSUBSGS), but going half a magnitude deeper.
% ---------------------------------------
%- NIRS0S (magnitude-limited):
%N=184 -> K-band observations: N=170 (one is AAT-image) (-> puuttuu
%184-170-1=13) -> K, or opt, or both: N=171
%
 %  -> mag-lim otoksen galaksit, joille ei K-havaintoja:
 %     eso137-34 - probably a distant galaxy with bright stars in the field (.SXS0?.)
 %     ic5250/50A - merger (.L...P?/.L...P?)
%      ngc147   -dE (.E.5.P.)
%      ngc185   - dE (.E.3.P.)
%      ngc205  - could have been observed making a mosaic (.E.5.P.)
%      ngc404  - bright star makes impossible to observe the galaxy (.LAS-*.)
%      ngc1291 - but we have SINGS image (RSBS0..)
%      ngc1316 - but we have SINGS image (.LXS0P.)
%      ngc1546 - simply not observed (.LA.+?.)	
%      ngc1808 - could have been observed making a mosaic (RSXS1..)
%      ngc1947 - simply not observed (.L..-P.)
%      ngc5128  - large nearby early-type with dust lanes (.L...P.)
% 
%      Kun poistetaan 1 kaukainen galaksi, 2 mergers, 2 dEs, 1 kirkas tahti kentassa,
%      2:lle SINGS, 1 suuri lahigalaksi, jaa 13-9 = 4 galaksia, jotka olisi voitu havaita, mutta johon
%      aika ei riittanyt.
%
%- all K-band observations: N=201 

The observations were carried out during the period 2003-2009 using
various ground-based telescopes in the two hemispheres, with sizes
between 2.5-4.2 m. The observing campaigns are shown in Table 1,
listing the pixel scale and field of view (FOV) of the
telescope/instrument setup used. The telescopes used were the 2.5m
Nordic Optical telescope (NOT, La Palma) using NOTCam, the 3.6m New
Technology Telescope (NTT, ESO) using SOFI, the 4.2m William Herschel
Telescope (WHT, La Palma) using LIRIS, the 3.6m Telescopio Nazionale
Galileo (TNG, La Palma) using NICS, the 2.1m telescope at Kitt Peak
National Observatory using Flamingos, and the 4m telescope at Cerro
Tololo Inter-American observatory (CTIO, Chile). Most of the galaxies fitted in the
typical 4-5 arcmin FOV, whereas for the largest galaxies the
19.5' FOV of Flamingos was used.  The total on-source integration time
was 1800-2400 sec, taken in exposures of 3-30 sec, depending on
galaxy brightness and telescope/instrument setup.  Owing to the high sky
brightness in the near-IR, and because the galaxies typically occupied
a large fraction of the FOV, an equal amount of time was spent
on the target and on the sky.  The target and the sky fields were
alternated after every 1-2 minutes using a dithering box of 20'' for
the target. Either sky or dome flatfields were obtained, depending
on what was recommended at each telescope.  The seeing conditions
were generally good (see Table 2), the full width at half maximum
(FWHM) being typically around 1''.  Seeing was worst at KPNO
(for 10 galaxies) where FWHM was between 2-3'', whereas at the NTT the FWHM
was below 1'' for most of the time (57 galaxies).  As the
flux calibrations were done using 2MASS images, flux calibration
standards were observed only occasionally.

In total, 206 galaxies were observed, including 172 galaxies of the
magnitude-limited sample of 185 galaxies. Of the non-observed 13 galaxies,
ESO 137-34 is most probably a distant galaxy having two bright stars
in the field. Two of the late-type ellipticals (NGC 147 and NGC 185)
appeared to be dwarf galaxies, and NGC 404 could not be observed due
to the bright star in the immediate vicinity of the galaxy. NGC 205,
NGC 1808 and NGC 5128 were too large to be observed with our typical
FOV, and at Kitt Peak these galaxies were not visible during the
period when time was allocated. IC 5250/5250A is an advanced merger
and therefore not useful for our analysis.  Four of the galaxies, NGC
1291, NGC 1316, NGC 1546 and NGC 1947, were not observed because of
a lack of observing time. However, for NGC 1291 and NGC 1316 SINGS
{\it Spitzer Space Telescope} images at 3.6 $\mu m$ are available (Kennicutt 2003).  In
conclusion, in our magnitude-limited sample there are only five galaxies of
interest (NGC 205, NGC 1808, NGC 5128, NGC 1546 and NGC 1947) for
which we lack NIR observations. Of these NGC 205 is a low surface
brightness galaxy, most probably an S0 with a central lens.  NGC
1808 is a dusty Sa-type spiral, whereas NGC 5128, NGC 1546 and NGC
1947 have strong dust lanes in a nearly featureless spheroidal
component, and are classified as $T$=$-$2, $-$1 and $-$3, respectively.  
Of the S0-S0/a galaxies in the magnitude-limited sample
observations for only four galaxies are missing, which means that
the completeness of our observations is 98$\%$.

\section{DATA REDUCTION}

\subsection{Combining the images}

The images were combined using IRAF routines \footnote{IRAF is
  distributed by the National Optical Astronomy Observatories, which
  are operated by AURA, Inc., under cooperative agreement with the
  National Science Foundation.}.  The main reduction steps consisted
of subtracting the sky from each science image, flatfielding the
difference image, combining the images after correcting the shifts
between the images, and fine-turning the sky subtraction.  The sky
images taken immediately before and after the target observation
generally worked best for the sky subtraction.  For flatfielding
normalized master flat-fields were used, made as an average of the
differences of high and low ADU-level images (ADU=digital counts).  In
the dome flats obtained at the NTT, scattered light sometimes produced
a shade pattern which was corrected using the correction frames
offered by ESO. While combining the images a 3 sigma clipping factor
was used to reduce the noise.  The images obtained at the WHT, TNG
and NTT showed ``crosstalk'', appearing as vertical or horizontal
stripes in the images. For the ES0/NTT images the script crosstalk.cl
(available at ESO) corrected the stripes effectively.  For the WHT
images this problem was more severe, and the stripes were corrected
manually using the IRAF routines IMCOPY and BACKGROUND. For some of
the galaxies bad lines/columns and sky gradients were also corrected.
Foreground stars were removed using the DAOPHOT package in IRAF, and
the cleaning was completed with the IMEDIT routine.  The images were
transposed to have north up and west to the right.

\subsection{Flux calibrations}

Flux calibrations were done using the $K_{\rm s}$ aperture photometry of
galaxies given in 2MASS\footnote{2MASS is a joint project of the
  University of Massachusetts and the Infrared Processing and Analysis
  Center/California Institute of Technology, funded by the National
  Aeronautics and Space Administration and the National Science
  Foundation.}. We write

  $$\mu = -2.5 \log_{10} F/pix^2 + \mu_0,$$

\noindent where $\mu$ is the surface brightness in units of
mag arcsec$^{-2}$, $F$ is the flux in digital units (normalized to
1 second), $pix$ is the pixel size in arcsecs, and $\mu_0$ is the magnitude zeropoint.  After sky
background subtraction and removal of foreground stars, the total flux
within a 14 arcsec (diameter) circular aperture around the galaxy center
was measured, and compared to the corresponding 2MASS aperture magnitude,
$m_{14}$, available via NED. The zeropoint $\mu_0$ was calculated from
the equation

$$\mu_0 = m_{14} + 2.5 \log_{10} (\sum_{r_i < 7''} F_i),$$

\noindent where  $r_{\rm i}$ is
the distance from the galaxy center.  In the calculation of
the total flux inside the aperture, bilinear interpolation was used for the
pixels falling on the aperture border. Also, the images were first
degraded to have the same seeing as the 2MASS images, to compensate
for the possible leaking of light in the original 2MASS aperture
measurements. We thus applied a convolution with a Gaussian PSF with

$${\rm {FWHM}_{conv}} = \sqrt{(2.5")^2 - {\rm {FWHM}^2}},$$

\noindent where FWHM corresponds to the original NIRS0S image, and $2.5"$
is the typical value for 2MASS images.  In practice, the
uncertainties in the photometric zeropoint due to sky subtraction,
centering of the aperture, or the applied bilinear interpolation are
all negligible ($<$0.001 mag).  Likewise, the effect of different
FWHM's is quite small ($<$0.02 mag), and thus the formal error of our
zeropoints corresponds to the accuracy of the 2MASS absolute
calibration, $\sim$ 2-3$\%$ (Jarrett et al. 2000).

As an additional check, and to minimize possible human errors (e.g.,
use of wrong image, wrong centering, pixel size etc.), we also used
the 2MASS $k_{20}$ and $k_{\rm ext}$ magnitudes to check the
consistency of our zeropoint calibration. These quantities were loaded
from NASA/IPAC infrared science archive (IRSA) via GATOR: $k_{20}$ is
the total magnitude inside $\mu_k$=20 mag isophotal ellipse, and
$k_{\rm ext}$ is the extrapolated total magnitude. IRSA lists the
isophotal radius $a_{k20}$, position angle $\phi_{k20}$ and axial
ratio $(b/a)_{k20}$, and the radius $r_{\rm ext}$ corresponding to
$k_{\rm ext}$ (isophotal orientation and shape are the same as for
$k_{20}$).

\begin{figure}
\includegraphics[width=140mm]{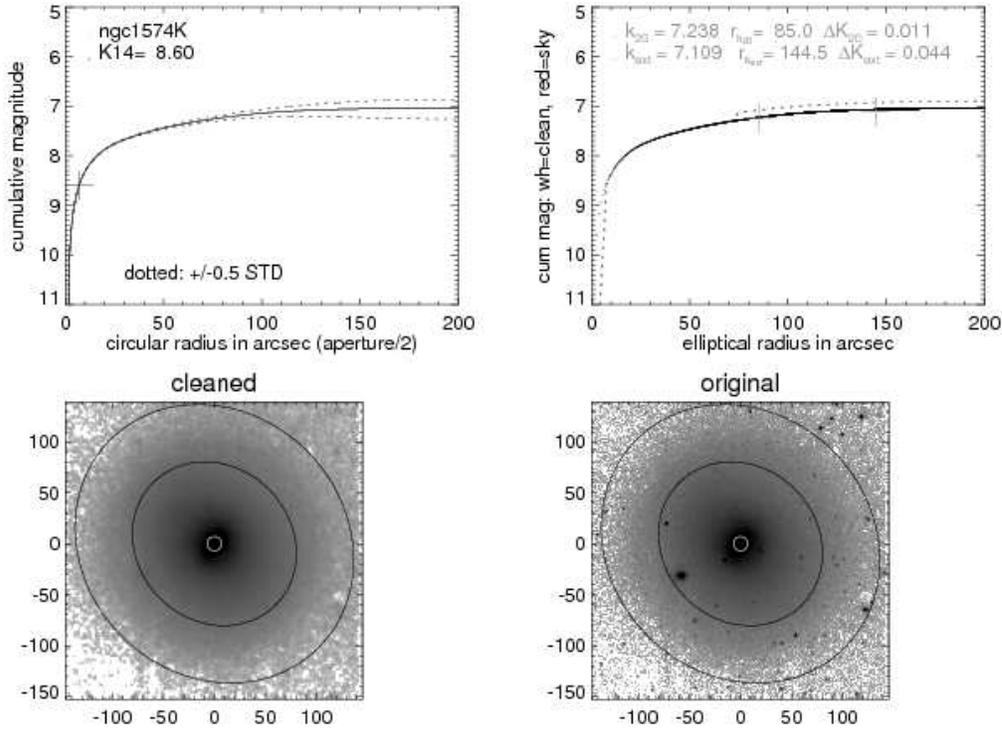}
\caption{Example of zeropoint calibration based on 2MASS 14'' circular
  aperture magnitude $m_{14}$. In the upper row, circular and
  elliptical isophote cumulative magnitudes are shown, while the lower
  row displays the original image (right) and the cleaned image
  convolved to FWHM=2.5'' (left). The circular aperture growth curve
is adjusted to go through $m_{14}$ at $r$=7''. The dotted lines
indicate the effect of adjusting the sky background by $\pm$ 0.5 times
the sky rms-variation, being completely negligible on the derived
$\mu_0$. The two crosses at the elliptical aperture growth mark the
2MASS $k_{20}$ and $k_{\rm ext}$, and their differences from the measured
curve are indicated. The black and red elliptical curve 
corresponds to cleaned and original images, illustrating the maximum
possible effect of star removal. The 2MASS 14 arcsec aperture (white)
and the used elliptical isophotes (black) are displayed on top of images.
(for all galaxies: http://www.oulu.fi/astronomy/NIRS0S$\_$pub/Kcalibration.html }
\label{figure-flux_cali}
\end{figure}

Figure \ref{figure-flux_cali} displays a typical example of flux
calibration (similar plots for all galaxies are available at the
NIRS0S website), displaying the cumulative magnitudes using both
circular (left) and elliptical (right) apertures. Also shown are the
NIRS0S images: on the left the cleaned image, convolved to
FWHM=2.5'', and in the right the original image before removal of
foreground stars. The elliptical aperture plot includes cumulative
magnitudes from both the cleaned (black line) and the original image (red
dashed line), to illustrate the possible effect of individual bright
stars. As described above, the circular aperture growth curve is
adjusted to go exactly through the $m_{14}$ point at $r$=7'', while
the elliptical aperture fluxes measured at $r_{k20}$ and $r_{\rm ext}$
usually deviate slightly from the tabulated $k_{20}$ and $k_{\rm ext}$.
We use the deviations of these quantities to control the possible
inaccuracy of the flux calibration, and list them in Table 2 for each
galaxy. Note that even large deviations do not indicate errors in our
calibration, based solely on the $m_{14}$ aperture magnitude: in some
cases the deviations are connected to bright stars near the galaxy, or
to the presence of a nearby galaxy. For a few cases, where
reliable background subtraction was difficult due to the small FOV
compared to the galaxy size, we made small adjustments to the sky
background based on matching the 2MASS $k_{20}$ value.

\begin{figure}
\includegraphics[width=140mm]{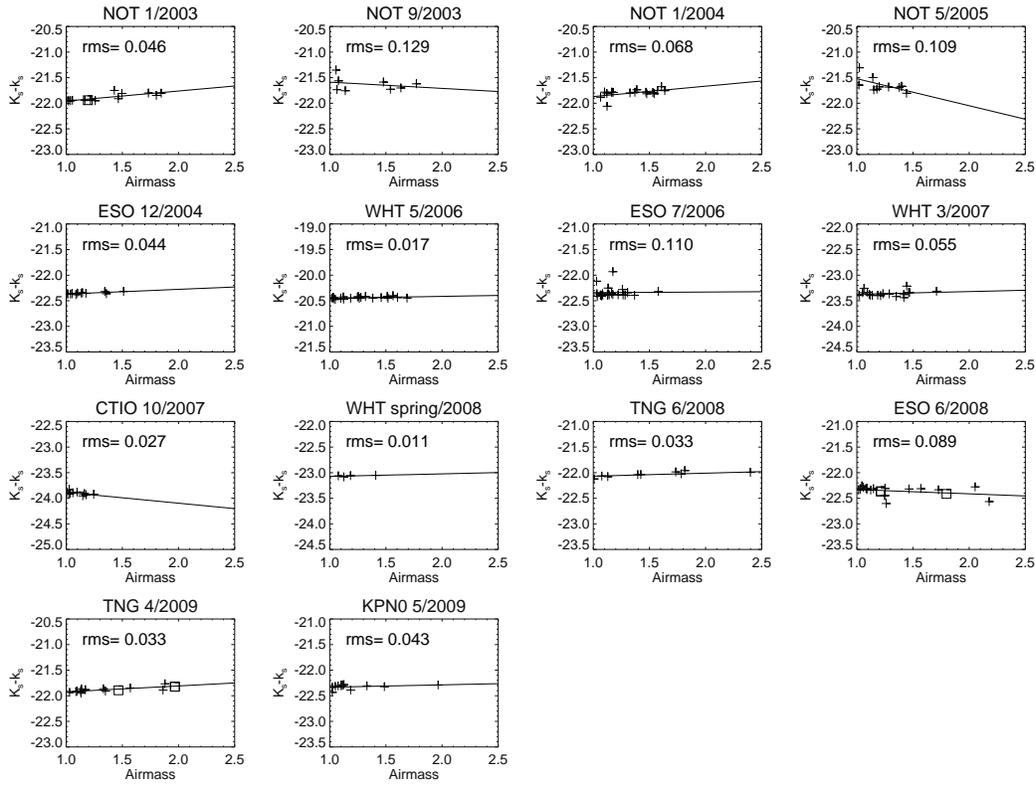}
\caption{The zeropoints derived based on 2MASS 14'' aperture
  calibration are displayed vs. airmass, for the 14 different
  observing campaigns. The rms scatter for each campaign is
  indicated. Boxes mark galaxies for which the zeropoint was adopted
  based on fitted campaign values, instead of using 2MASS aperture measurement.}
\label{figure-kampanja}
\end{figure}

In case 2MASS data was not available (NGC 1161), or when there were reasons
to believe that the above calibration is not reliable (interacting
systems), we adopted the average zeropoint value derived for the
NIRS0S observing run in question (see Fig. \ref{figure-kampanja}). For
some interacting pairs the difference in $\mu_0$, compared to that
obtained directly from the 2MASS $m_{14}$ calibration, is less than 0.1
magnitudes (NGC 2292/2293, NGC 4105/4106), but in some cases the
difference is larger (NGC 5353/5354, NGC 5636, NGC 6438).  For
NGC 4474, for which the 2MASS values $m_{14}$, $k_{20}$, and $k_{\rm ext}$
are mutually inconsistent, the 2MASS calibration was not used.

For comparison, standard stars were observed during
one campaign, at the NTT in 2004.  Flux calibration standards of
\citet{persson1998} were used: 10 standards at each night were
observed, using an observing block where the star was integrated in
the four corners of the  frame. The images were combined in a
similar manner as the science images.  Using the standard star
measurements the following zero-points and extinction coefficients
were obtained for the three nights:

$K_{\rm s}$ = $k_{\rm s}$ + 22.399$\pm$0.072 - 0.062$\pm$0.061 $X$ (first night)

$K_{\rm s}$ = $k_{\rm s}$ + 22.350$\pm$0.036 - 0.020$\pm$0.029 $X$ (second night)

\begin{figure}
\includegraphics[width=140mm]{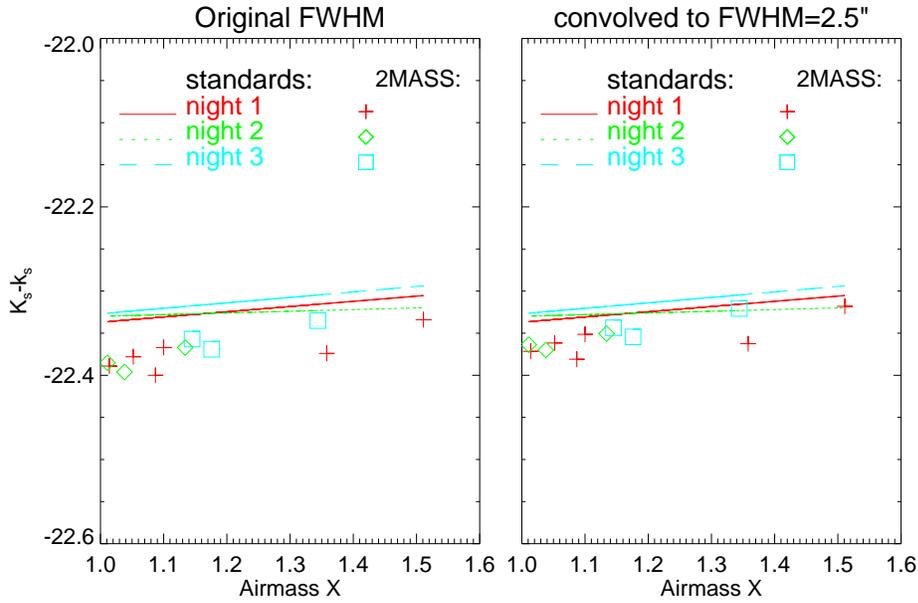}
\caption{Comparison of 2MASS and standard star calibration based
  zeropoints ($\mu_0 = -(K_{\rm s} -k_{\rm s})$). The three curves with different
colors indicate linear fits for zeropoint versus airmass, obtained by
observing 10 standard stars per night. The symbols indicate zeropoints
derived for the galaxies observed during the same nights, based on
2MASS calibrations. In the left, applying 2MASS calibration to
original images (with typical $FWHM$ $\sim$ 1-2''), there is about 0.04
mag shift between the calibration methods. However, after allowing for
the poorer seeing of 2MASS images (FWHM = 2.5'') the systematic shift
is about 0.02 mag.}
\label{figure-standard_star}
\end{figure}

$K_{\rm s}$ = $k_{\rm s}$ + 22.392$\pm$0.041 - 0.065$\pm$0.03 $X$ (third night),

\noindent where $K_{\rm s}$ is the total magnitude in the photometric
system, $k_{\rm s} = -2.5 \log_{10} (\sum F_{i,star})$ is the
instrumental magnitude of the star, and $X$ is the airmass. The
zero-points and extinction coefficients are very similar for the first
and the third nights, which were photometric.  Comparison with the
2MASS-based calibration (Fig. \ref{figure-standard_star}) shows very
good agreement: at most there is a marginal 0.02 magnitude systematic
shift, which, however, is comparable to the internal scatter of the two
sets of calibrations. Without the convolution to FWHM = 2.5'',
the systematic difference would be clear, about 0.04 mags.  Based on
this observing campaign, we estimate that any possible systematic
error in $\mu_0$ introduced by using the 2MASS-based calibration is
less than 0.05 magnitudes.

\begin{figure}
\includegraphics[width=140mm]{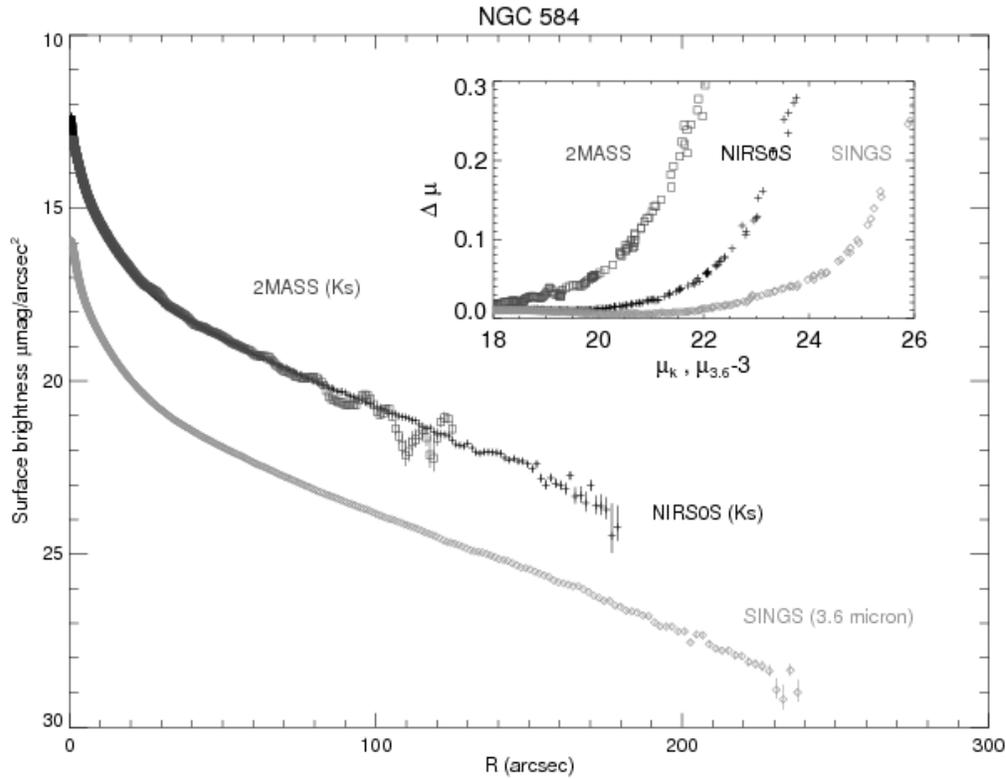}
\caption{Example of NIRS0S surface brightness profile for NGC 584. The
  large figure shows the brightness profile vs isophotal radius,
  obtained with IRAF ellipse routine. Fixed orientation and
  ellipticity are used: $PA$=73$^o$.6 and $q$=0.675 correspond to estimated outer
  disk orientation. Error bars indicate the uncertainties $\delta
  \mu$ returned by ellipse-routine. Before calculating the profile,
  the NIRS0S image was rebinned by a factor of 3, to pixel size
  0.86'' For comparison, also the profile derived from 2MASS Atlas
  image (pixel size 1''), and Spitzer SINGS survey (pixel size 0.75'',
  3.6 micron IRAC1 band) are shown: allowing for the $\approx$ 3 mag
  shift between the $K_{\rm s}$ band and the $3.6 m \mu$ $AB$-system
  magnitudes, all profiles agree well, except for extending to
  different depths. The inserted figure shows the $\Delta \mu$ vs $\mu$
  (taking into account the aforementioned difference between
  2MASS/NIRS0S and SINGS magnitude systems), illustrating the $\approx$
  2-3 mag differences in depth between the images.}
\label{figure-depth}
\end{figure}

The NIRS0S images are deep: Table 2 lists the 
1-sigma sky deviation per square arcsec, calculated from

$$\sigma_{\rm sky} = -2.5 \log_{10} (\Delta F_{\rm sky}/pix) +\mu_0,$$

\noindent where $\Delta F_{\rm sky}$ is the sky rms-variation,
obtained by measuring it at several locations outside the
galaxy.  Depending on the telescope, exposure time and sky conditions,
the values range from 21 to 23 mag arcsec$^{-2}$, with mean
$<\sigma_{\rm sky}> \approx$ 22.2 mag arcsec$^{-2}$.  However, the
azimuthally averaged surface brightness profiles are more illustrative
than $\sigma_{\rm sky}$ to show the real image depth, extending to
23-24 mag arcsec$^{-2}$, depending on the galaxy.  A typical example is
shown in Figure \ref{figure-depth} for NGC 584 (with $\sigma_{\rm
  sky}$=21.8 mag arcsec$^{-2}$), with profiles from the 2MASS Atlas
image (Jarrett et al. 2000) and from the SINGS survey 3.6 micron image
(Kennicutt 2003) overlaid for comparison. For NGC 584 the useful NIRS0S profile
($\Delta \mu \le$ 0.2) extends to about 23.5 mag arcsec$^{-2}$, or 2-3 magnitudes deeper
than 2MASS. Allowing for the difference in the band and magnitude
system, it is only about 2 mag shallower than the deep {\it Spitzer}
image (see the insert in Figure \ref{figure-depth}).
%For the best NIRS0S images (extending to depths
%$\sim$ 25 mag/arcsec$^2$ in Ks, see Fig. \ref{figure-moredepth}) the
%difference to corresponding Spitzer images would be less than one
%magnitude.
In $B$-magnitudes, 23.5 mag arcsec$^{-2}$ in the $K_{\rm s}$-band translates roughly to a surface
brightness of 27.5 mag arcsec$^{-2}$. However, not all of the galaxies
in our sample are visible at this surface brightness level, for-instance
because the FOV is too small, the galaxies are strongly interacting,
or in a very few cases because the sky background is not stable enough,
in which case the sky gradients limit the useful image depth.
In radial extent our example galaxy is 1.4-1.6 times larger than the
2MASS image, and 0.8 times of the extent of the SINGS image. 

\section{THE IMAGE ATLAS}

\subsection{The Atlas images}

The flux-calibrated image Atlas is shown in Figure 5. Examples of
  atlas images are shown for the galaxies discussed in the text,
  whereas the complete Atlas is available in Supporting information with the
online version of the article. 
% http://www.oulu.fi/astronomy/NIRS0S\_pub/atlas.html; for .fits images
%contact PI of this paper). 
The images are
  shown in logarithmic form, in units of mag arcsec$^{-2}$,
  maintaining the full pixel resolution (upper panel).  The magnitude
  range is given in the right hand bar, which is selected individually
  for each galaxy so that the full scale of structures is visible.  In
  order not to add any artifacts due to bad foreground star removal,
  the images before star removal are shown.  Our visual and
  photometric (in brackets) classifications (see Section 5) are marked
  in the figures.  A drawback of this layout is that it does not give
  full justice to the real image depth, failing to show the faint
  outer regions.

These faint outer structures are shown in one of the small panels, where
rebinned images cleaned of foreground stars are shown.  The galaxies are
shown in many different radial and brightness scales, the number of
frames depending on the complexity of a galaxy's morphology.  In some
of the galaxies faint bars, lenses or dust lanes are overshadowed by
prominent bulges. These components were made visible using 2D
multi-component structural decompositions previously given for the
Atlas galaxies by Laurikainen et al. (2010): the bulge model was
subtracted from the original image leaving the faint structures
visible in the residual images. Alternatively, unsharp masked images are
shown. They were created by smoothing the images by 5-20 pixels,
and then subtracting smoothed images from the original
images. The upper left small panel shows the image rebinned by a factor that
best demonstrates the faint outer structure of the galaxy, whereas the
lower right panel generally shows either the residual image or the
unsharp masked image. In such cases a text ``bsub'' or ``unsharp'' is
overlayed on the image. The detected faint structures form part of our
photometric classification.
%        classification E, or E/S0:
%
%       IC1392, IC4889, IC4991, ESO208-21, ESO337-10,
%       NGC1400, NGC2292(ia),
%       NGC3665, NGC4073, NGC4105/6, NGC4281, NGC4373, NGC4435, NGC4459,
%       NGC4638, NGC4649, NGC4976, NGC5087, NGC5078, NGC5266, NGC5353, 
%       NGC5485, NGC5493, NGC6407, NGC6482, NGC6861, NGC7029,
%       NGC7049  
%
%Fig. 3: Atlas of K-images (mag-scale, bar showing the mag-values)
%
%Fig. 4: The Atlas galaxies shown in small panels (show sub-structures) + ellips%e fitting plots
%
%http://www.oulu.fi/astronomy/nirs0s/ATLAS_101210_index.html - NIRS0S Atlas

\subsection{Ellipse fitting}

The atlas figures show also isophotal analysis results, which consist of
fitting elliptical isophotes to the images using the {\it ellipse}
routine in IRAF. This routine uses a technique in which Fourier series
are fitted to concentric isophotes \citep{Jedr1987}. The quality of the
fit is evaluated by inspecting the one-dimensional brightness
distribution as a function of position angle, so that the harmonic
content of this distribution is analyzed. The fourth order coefficient
$b4$ of the best fit Fourier series then measures the isophote's
deviations from perfect ellipticity.  We calculate the radial profiles
of the position angle PA, the ellipticity $\epsilon$ = 1-$q$ (where $q$=$b/a$
is the minor-to-major axis ratio), and the parameter $b4$.
Non-rebinned images were used and the center was fixed to the value
estimated by the IMCNTR routine in IRAF.  Logarithmic radial spacing
was used along semi-major axis while fitting the elliptical
isophotes.  In order to minimize the effects of noise and
contamination by bad pixels and cosmic rays, deviant pixels above 3
$\sigma$ were rejected.

The surface brightness profiles and the radial profiles of PA,
$\epsilon$ and $b4$ are shown in Figure 5. The full green vertical
line shows $r_{20}$, which is the 2MASS isophotal radius of the 20 mag
arcsec$^{-2}$ contour in $K_{\rm s}$-band.  The radius
$r_{20}$ is marked with a black ellipse in the upper left panel, using
2MASS position angle $\phi_{k20}$ and axis ratio $(b/a)_{k20}$. The
dashed vertical lines show the radii of the bars in our classification
(see Section 5).  In some cases the $r_{20}$ isophotal orientations
from 2MASS deviate from the orientations in our images. This is
because for these galaxies the 2MASS images trace only the inner
components of the galaxies, the outermost components detected by us
having different orientations.

\section{MORPHOLOGICAL CLASSIFICATION}

\subsection{Brief history}

The identification of S0s traces back to \citet{lundmark1926} and
\citet{reynolds1927}, who recognized a group of amorphous galaxies
without any sign of spiral arms, a group of galaxies which form the
modern class of E+S0 galaxies.  Early-type galaxies were seen as a sequence of
increasing flattening towards later types. In {\it Realm of the Nebulae}
Hubble (1936) described hypothetical S0s which, based on Hubble's
notes, Sandage (1961) included in the Hubble sequence as {\it
  transition types between the ellipticals and the spirals}. As a real
physical scenario this idea was abandoned when it was found that S0s
have a similar flattening distribution as spirals, which clearly
deviates from that for the elliptical galaxies \citep{sandage1970}.

De Vaucouleurs (1959) further refined the Hubble/Sandage
classification by adding the {\it stage} (S0$^-$, S0$^0$, S0$^+$),
{\it family} (A, AB, B), and {\it variety} (s, r), as well as the
outer ring/pseudoring designation to the S0 class. Concerning spiral
arms, he was less restrictive in the sense that the (r) and (s)
varieties were carried even into the earliest S0 classification,
called S0$^-$, although such cases are difficult to recognize.  In the
classification by \citet{sandage1981} the stage was given in a similar
manner as in de Vaucouleurs' classification, only different symbols
were used (S0$_1$, S0$_2$, S0$_3$). They also added flattening of a
galaxy into their coding. Outer ring classification was fine-tuned by
\citet{buta1991} and \citet{buta1995}.  The morphology of bars in
terms of boxy/peanut/x-shaped structures were included in the
classification by \citet{butaetal2010} for 200 galaxies using Spitzer
mid-infrared images (S$^4$G; Sheth et al. 2010). Buta et al.  also
suggested a notation 'nb' for nuclear bars, and 'nr' for nuclear
rings.  Kinematic observations for edge-on galaxies \citep{kuijken,bureau1999} have shown
that boxy/peanut/x-shaped structures are inner parts of bars.
Although lenses form part of the classification of
S0s they were not coded into the morphological classification.  For
lenses, \citet{kormendy1979} suggested a coding where 'l' stands for
an inner, and 'L' for an outer lens (as used in the dVA), whereas
\citet{butaetal2010} suggested a notion 'nl' for nuclear lenses.

The classification has been developed along with new ideas of galaxy
formation and evolution. That was already the case when
\citet{baade1963} suggested that S0s are stripped spirals formed in
galaxy interactions. The idea was developed by \citet{berg1976}
leading to a classification where the S0s form a sequence from early-
to late-types, similar to that used for the spirals (S0$_a$, S0$_b$,
S0$_c$).  It was based on the supposition that there may exist anemic
S0s, which have similar surface brightness distributions as the Sa, Sb
and Sc-type spirals.  In this scenario, also small low luminosity
bulges are expected among S0s.  The hypothesis was tested by
\citet{sandage1994} for 200 bright S0s, but no such S0s were found.
However, multi-component structure analysis has cast new light on
this approach (see Laurikainen et al. 2010). A dust-penetrated
classification was suggested by \citet{buta2001}, where a
bar-induced torque forms part of the classification. This was based on
the idea that bars are a driving force of secular evolution, thus
modifying galaxy morphology.

Early-type disk galaxies typically have faint structures that are
easily missed in visual classification. In particular, 
late-type ellipticals and early-type S0s are difficult to
distinguish due to the subtle oval or twisted structures, or due to
faint outer disks in S0s. Indeed, \citet{sandage1994} report many
misclassified S0s in the RC3. Therefore, sometimes photometric
classification is also used. For example, \citet{kormendy2009} used
isophotal analysis to show the appearance of disks in galaxies in which the
disks were not obvious in the direct images. It was assumed
that disks are more flattened than bulges.

In the current study, both visual and photometric classification is made,
given in Tables 3 and 4. For comparison, in Table 3 we give also the
classifications from the RC3 and the RSA.  Our classification is purely
morphological, and no assumptions of possible formative processes of
galaxies are made.  Although our images do not have the resolution of
the {\it Hubble Space Telescope}, they are still good enough for detecting also
nuclear bars, rings and lenses.

\subsection{Visual classification}

We use classification, based on de Vaucouleurs' revised Hubble-Sandage
system \citep{vaucouleurs1959} (see also the dVA and Buta 2011), which
includes the {\it stage} (S0$^-$, S0$^o$, S0$^+$, Sa), the {\it
  family} (SA, SAB, SB), the {\it variety} (r, rs, s), the {\it outer
  ring or pseudoring} (R, R'), possible {\it spindle} shape (sp,
meaning edge-on or near edge-on orientation), and the presence of {\it
  peculiarity} (pec).  We use also a notation for shells and ripples.
The underline notation (e.g., S$\underline{\rm A}$B, SA$\underline{\rm
  B}$,$\underline{\rm r}$s, r$\underline{\rm s}$) as used by
\citet{vaucouleurs1963} is used to emphasize the more likely
phenomenon in a galaxy. Notice that in the atlas images (in Fig. 5),
for technical reasons, the underline notation is emphasized by slanted
font instead.  Following \citet{buta1991} and \citet{buta1995} we also
recognize subcategories of pseudorings (R$_1$, R$_1'$, R$_2'$,
R$_1$R$_2'$).  Representative examples of {\it stage} and {\it family}
for barred and non-barred atlas galaxies are shown in Figure 6.  De
Vaucouleurs' classification does not include the morphology of bars in
terms of boxy/peanut/x-shaped structures (B$_x$), which is included in
our classification. Bars can also have classical rectangular
structures, or ansae at the two ends of the bar
\citep{lauri2007,valpuesta2007}, which ansae types are coded by B$_a$
in our classification. Examples of bar and ring morphologies in S0s
are shown in Figure 7. It is worth noticing that the x-shaped
structures inside the bars in our study, and in dVA, appear in fairly
face-on galaxies, not in edge-on systems where they are generally
reported (see for example IC 5240 and NGC 4429 in Fig. 5).

Nuclear, inner and outer lenses are denoted by 'nl', 'l' and 'L',
respectively. Nuclear bars and nuclear rings, denoted as 'nb' and
'nr', have similar sizes as nuclear lenses.  In the classification,
intermediate types between rings and lenses are also used (nrl, rl,
RL).  Additionally, a new lens type is introduced which we propose to
call ``barlenses" with a notion of 'bl'. These appear in the central
regions of many NIRS0S galaxies, but are generally distinct from
nuclear lenses by their much larger sizes.  From visual appearance
these 'bl' features can be mistaken for large bulges.  The appearance
of 'bl' is demonstrated for NGC 2983 in Figure 8, where both the
original and the residual images are shown: the residual image is
created by subtracting from the original image the bulge+bar model
obtained from 2D structural decomposition.  A manifestation of 'bl'
appears also in NGC 4314 (Fig. 9): the fine-structure in the central
regions confirms that the component cannot be a bulge. Also, as the
galaxy is in nearly face-on orientation the fat bar component cannot
be interpreted as a boxy bar structure seen nearly edge-on
orientation. This needs to be explained by the theoretical models, in
which the boxy/peanut structures are generally induced by vertical
thickening of the bar (Athanassoula $\&$ Misioritis 2002).

Due to our selection criteria the sample should not contain edge-on
galaxies.  However, several misclassified galaxies appear in the
RC3. We have moved the following galaxies in our sample to the spindle
category: ESO 208-21, IC 1392, NGC 2685, NGC 3414, NGC 4220, NGC 4435,
NGC 4474, NGC 4546, NGC 5087 NGC 6861 and NGC 7029. On the other hand,
the galaxies NGC 4281 and NGC 5353 which were classified as spindle in
the RC3, were considered as more face-on systems in this work. Notice
that the galaxies NGC 4638, NGC 5493 and NGC 7029 have an edge-on disk
clearly shorter than the outskirts of the IR spheroidal components,
which have boxy outer isophotes (Fig. 5) . We use the notation of Kormendy and
Bender (1996) for the boxy elliptical parts of these galaxies with
bright embedded S0 disks, although these galaxies do not fit very well 
to the scheme by Kormendy and Bender.

Our classification is similar to that used by \citet{butaetal2010},
except that lenses are coded in a systematic manner in the present
work.  We have 12 galaxies in common with that sample. As expected,
the agreement is generally good, except that in six of the galaxies we
detect lenses (NGC 4203, Fig. 14; NGC 4245, Fig. 5; NGC 4314, Fig. 9;
NGC 4649, Fig. 5; NGC 5377, Fig. 7; NGC 5846, Fig. 5), which were
not recognized by Buta et al. Also, for NGC 5353 (Fig. 5) we recognize an
x-shaped bar which was not included in the classification by Buta et
al.  On the other hand, for NGC 4369 (Fig. 5) Buta et al. detect an outer ring,
which we don't see in the $K_{\rm s}$-band image, most probably
because the image used by us is not as deep as the {\it Spitzer} image
at 3.6 $m \mu$ of Buta et al.

Compared with the optical classification in the RC3, the Hubble stage
differs for some individual galaxies.  However, deviations appear in
both directions, so that there is no systematic shift in the mean
Hubble stage ($<T>$=$-$1.57 and $-$1.51 in near-IR and optical,
respectively). The scatter plot of the optical and NIR-types is shown in Fig. 10.
Both \citet{eskridge2002} and \citet{butaetal2010}
found that intermediate-type (S0/a-Sc) galaxies are on average one stage earlier in the
infrared than in the optical. The reason why we don't see such a shift
is partly because we study early-type galaxies which have only a small
amount of dust, whereas the samples by Eskridge et al. and Buta et
al. are more concentrated on dusty spirals. Another reason is that,
although some of the galaxies in our sample were shifted towards an
earlier stage, that is partly compensated by shifting some ellipticals
that were misclassified in the RC3 into S0s in our classification.

\subsection{Photometric classification}

By photometric classification we mean including faint structure
components, even if they were not obvious in visual classification,
for example because they were outshone by luminous bulges. Also,
galaxies that are late-type ellipticals (E$^+$) 
in visual classification, can
turn out to be early-type S0s in photometric classification if
exponential outer disks are detected from surface brightness profiles.
Our classification is based on morphology alone, and does not include
any kinematic observations or parameters measured from the images
(ellipticities, bulge-to-total flux ratios, or bar strengths). Also,
no assumptions on the formative processes of galaxies were made,
for example by assigning features like shells/ripples, assumed to be
merger-built structures, to elliptical galaxies alone. The photometric
classification is given only if it deviates from the visual
classification.

\subsubsection{Subtle features not identified visually}

In order to identify faint structures we use (1) unsharp masks
(see Section 4.1), or (2) residuals from structural decompositions. We
use 2D multi-component decompositions obtained previously for the
Atlas galaxies by Laurikainen et al. (2005, 2006, 2010).  The
decompositions were made by fitting a S\'ersic function for the bulge,
an exponential function for the disk, and a Ferrers function for the
bar. In some of the galaxies more than one bar was fitted. Lenses
were fitted either by Ferrers or S\'ersic functions. Faint
structures are visible in the residual images after subtracting a
bulge model. Such structures can be bars, rings, inner disks or
dust lanes. Lenses were identified directly from the images or from
the surface brightness profiles: they were included to the parameter
fitting of the structure components and therefore generally do not
appear in the residual images.

The identification of lenses is an important part of our
classification, and the main lens types are shown in Figure 11. For the
galaxies in that figure the structural decompositions are also shown,
taken from \citet{lauri2010}.  Prominent lenses, like the one in NGC
2902, are directly visible in the images. Faint or very extended
lenses and ring/lens systems may not be immediately obvious in the
images, but can be identified as broad bumps or exponential
subsections in the surface brightness profiles (e.g., Laurikainen et
al. 2010).

NGC 1533 (Fig. 11) is a good example of a galaxy having an outer
ring/lens structure.  In the surface brightness profile it is
manifested as a S\'ersic profile with $n$ parameter smaller than one.
In NGC 2902 (Figs. 6, 11), the inner ring/lens (rl) feature is
manifested in a similar manner, causing a bump in the surface
brightness profile.  In NGC 524 (see Figs. 11 and 15), the lenses are
weaker and appear as exponential subsections in the surface brightness
profile. NGC 524 is one of the best examples of a largely face-on
(L)SA(l,nl) system, others being NGC 5846 (Fig. 5) and 5898 (see
Fig. 5), with related examples being NGC 1411 (see Fig. 16) and 7192
(Fig. 15). Our image of NGC 1411 is not deep enough to detect the
outer lens, seen in the dVA image in the optical
region.  In the decomposition of NGC 524, the inner (l) and nuclear
(nl) lenses are fitted by Ferrers functions. NGC 2983 
(Figs. 11, 14, 16) is an example of a barred galaxy having an outer lens
(L), and also a barlens (bl).
 
In Figure 12, images and structural decompositions are shown for
NGC 4459 and NGC 4696.  In visual classification these galaxies are
ellipticals, but the surface brightness profiles show perfect exponential shapes,
which changes the classification into S0. For these galaxies we have also
$B-K$ colour maps (H. Salo et al. 2011, in preparation), where the
lenses can be identified as colour changes in the interface regions
between lenses and disks.

In our classification we do not code ovals as a distinct morphological
feature, because they are often difficult to distinguish from
lenses.  Ovals are global deviations from an axisymmetric shape in
galactic disks (see Kormendy $\&$ Kennicutt 2004 and Buta 2011). In isophotal and
Fourier analysis they appear in a similar manner as bars (with 
higher Fourier modes, see Laurikainen et al. 2007), but with lower
ellipticities. In contrast to lenses, they have less shallow surface
brightness distributions. 

\subsubsection{Comparison with \citet{kormendy2009}}

We have five galaxies in common with the sample of
\citet{kormendy2009}, who studied mainly ellipticals and Sph type
spheroidals, but whose sample also includes some bright S0s. The
common galaxies are NGC 4382, 4472, 4552 and 4649 (Fig. 5), and NGC
4459 (Fig. 12), which were classified as ellipticals (mainly E2) by
Kormendy et al., and as S0s by us. They are S0s also in the
classification by Sandage $\&$ Tammann (1981).  For the last four
galaxies the decompositions by Laurikainen et al. (2010) have shown
that the galaxies can be fitted by a S\'ersic bulge and an exponential
disk. In NGC 4649 a lens was also identified by us.  Except for NGC
4459, Kormendy et al. report these galaxies as ellipticals that miss
light in the nuclear regions.

NGC 4382 is peculiar and therefore difficult to classify. Kormendy et
al. showed that the galaxy has extra light at intermediate radii above
a single S\'ersic fit, and also distorted isophotes. However, in their
view the extra light cannot be associated with a large-scale disk
as is typical for S0s.  The galaxy has shells/ripples, which is why
Kormendy et al. interpreted it as a merger remnant that has not yet
fully settled into an equilibrium.  The reason why we consider it to be
a disk galaxy and not an elliptical, is the detection of dispersed
spiral arm segments, but the exponential nature of the outer profile is not
clear. However, there are other S0s with shells/ripples which do have
clear extended exponential disks. Such galaxies are, for example, NGC
2782 and NGC 7585 (Fig. 5).

\section{DIMENSIONS OF THE STRUCTURES}

The dimensions, orientations (PA), and minor-to-major axis ratios (q)
of the structure components in our classification were measured, and
are given for rings and lenses in Table 5, and for bars in Table
6. The dimensions of the structure components are semi-major axis
lengths.  To measure the rings and lenses the following strategy was
used: after displaying the image rebinned by a factor of two, the
classified features were mapped visually, at least three times in
succession.  If a feature is a clear ring, the cursor was used with
IRAF routine TVMARK to outline the ridge-line. If the feature is a
lens, oval, or a bar, the edge was mapped instead.  After obtaining
x,y coordinates of the feature's location mapped in azimuth, an
ellipse-fitting program was used to fit the points to get the central
coordinates, the position angle of the major axis, the major and minor
axis radii, and the axis ratio. As an illustration of our strategy the
fitted ellipses for four features superimposed on the galaxy image are
shown for NGC 1543 in Figure 13. For this particular galaxy the
nuclear bar is also shown. Similar figures for the complete sample are
available in electronic form  (address given by MNRAS; http://cdsarc.u-strasbg.fr/cats).
%({\tt http://www.oulu.fi/astronomy/NIRS0S\_pub/nirsos\_dimensions.html}).

For measuring bar lengths three methods were used: (1) they were
estimated visually by marking the outskirts of the bar and drawing an 
ellipsoid to that distance ($r_{\rm vis}$ in Table 6). A line was also
drawn along the bar major axis which gave a visual estimate of the bar
orientation.  (2) Radial profiles of the ellipticities were used: bar
length was taken to be the radial distance where the maximum
ellipticity in the bar region appeared (following, e.g., Wozniak $\&$ Pierce 1991;
Wozniak et al. 1995; $r_{\rm ell}$ in Table 6).  (3) As a third
estimate, the bar length was taken to be (Erwin $\&$ Sparke 2003):

  $r_{\rm L}$ =  $r_{\rm ell}$ + ($r_{\rm ellmin}$ - $r_{\rm ell}$)/2,

\noindent where $r_{\rm ellmin}$ is the radial distance where the
first minimum appears after the ellipticity maximum in the bar
region. In galaxies with complicated morphological structures no
minimum appears after the maximum ellipticity; in those cases no
$r_{\rm L}$ is given.  $r_{\rm ell}$ is not given if the ellipticity
maximum was very broad, and also when the bar orientation in respect
of the disk orientation was not favorable.  

This is the case, for example for NGC 3384 (Figs. 5 and 13 in
electronic form), having a bar perpendicular to the disk which bar is
also inside a lens. As a consequence there is a minimum in the
ellipticity profile and in the $b4$-profile at the edge of the bar,
and a maximum in the position angle (this also means that the method
based on detecting maxima in the ellipticity profile miss bars in
unfavorable orientation).  An other similar case is NGC 4546 (Fig. 5,
electronic Fig. 13). If the bars are very weak and appear only in the
unsharp masks or in the residual images, only visual estimation of the
length can be given.  Typically $r_{\rm vis}$ is close to $r_{\rm
  ell}$, whereas $r_{\rm L}$ gives an upper limit for bar length.  The
standard deviations given in column 7 of Table 6 were calculated for
$r_{\rm vis}$ and $r_{\rm ell}$. The bar orientation is generally the
position angle near the ellipticity maximum in the bar region, but for
very weak bars, bars seen inside prominent lenses, or in unfavorable
viewing angle, the position angle from ellipse fitting could not be
used. For those galaxies visually estimated bar orientation is given.
Visually estimated orientation and those obtained from ellipse fitting
generally agree well, the variations typically being around 2 degrees.

\section{DISCUSSION}

It has been suggested that galactic disks are primary components of
galaxy formation, and that their surface brightness distribution
reflects the specific angular momentum distribution of protogalaxies
(e.g. Fall $\&$ Efstathiou 1980; Dutton $\&$ van den Bosch 2009).
There can also be angular momentum exchange between material at
different radii, in which case the exponential surface brightness
distribution is a result of disk viscosity
\citep{lin1987}. \citet{bosma1983} suggested that lenses might be
primary components formed soon after the disk formation: the outer
edges of lenses may have formed when the initial amount of gas
suddenly dropped and star formation abruptly ceased. Lenses may also
form by disk instabilities in a similar manner as bars
\citep{atha1983}.

On the other hand, in the hierarchical picture of galaxy formation
present-day galactic disks are merger-built structures, which have
been significantly restructured in galaxy collisions
\citep{white1978,kauffman1999}.  It has been suggested that even up to
$\sim$ 50$\%$ of all spiral disks might come from disk rebuilding from
recent mergers, the other half of the disks being formed in some earlier
mergers \citep{hammer2009}. If S0s are descendants of these spirals,
it needs to be understood how the multi-component bar/lens structures
that we find in up to 25$\%$ of the S0s were formed and maintained.
The statistics and a more thorough discussion of possible formative
processes of lenses will appear in forthcoming papers. Here we give
only tentative examples of possible morphological sequences of lens
formation, and discuss possible candidates of S0$_c$ galaxies.
 
\subsection{Morphological evidence of lens formation?}

1. {\it Lenses might be highly evolved star forming zones or highly evolved
stellar rings} (Fig. 14a). The upper panels show the full images, whereas
the lower panels show the inner regions of the residual images,
obtained after subtracting the bulge models from the original images,
for NGC 3998 and NGC 4203. In this scenario, gas is used by star
formation in the disk, leading to a dynamical heating so that the
spiral arms disappear, first in the outer disk (NGC 7371). However, in
the presence of a weak bar some of the material in the spiral arms may
be trapped into the resonances of the bar before all the gas is
consumed. Consequently, a ring or a double ring may form outside the
bar (NGC 3998). When the rest of the gas is consumed and the
inner disk is dynamically heated, the rings are expected to lose their
identity and a lens forms (NGC 4203).  Notice that in NGC 4203 the
lens clearly has a larger radius than the bar.  In \citet{lauri2009},
NGC3998 was interpreted as a possible candidate of bar destruction in a
spiral galaxy that was formerly barred.
%****Are we also arguing here that some lenses are merely highly evolved 
%former star-forming zones? Could some lenses be highly evolved stellar
%rings?****

2. The sequence NGC 5953 -$>$ NGC 7742 -$>$ NGC 7213 (Fig. 14b) is 
{\it similar to the sequence above, except that the galaxies have no bars}.
NGC 5953 has prominent flocculent spiral arms in the inner part of the
disk. If there is enough gas in these spiral arms a starburst may occur,
leading to a rapid increase in the stellar mass, which may take the form of
a lens with the dimension of the current extension of the spiral arms.
As an intermediate stage the spirals may take the form of rings, which is
clear in NGC 7742 (the ring in this galaxy is counter-rotating; de Zeeuw et al.
2002), and to some extent also in NGC 7213. 

3. {\it Lenses might be triggered by bars}, which is illustrated in
Figure 14c.  Bars are known to excite resonance rings, which can be
full classical rings, or of R$^{\prime}$ type rings, of which NGC 6654 is a good
example. When the gas is consumed the disk is
dynamically heated, and the ring in NGC 6654 may evolve into a lens,
similar to that surrounding the bar of NGC 2983. An alternative
progenitor type of NGC 2983 could be NGC 1326 (see Fig. 7), which has a prominent
dispersed ring surrounding the bar. When evolved over time the ring
may evolve into a lens. 

Several stripping mechanisms in spiral galaxies are suggested
  leading to significant decay of star formation and heating of the
  disk, which is required while transferring spiral galaxies into
  S0s. According to numerical simulations (Bekki $\&$ Couch 2011)
  reduced star formation is particularly important among barred
  interacting galaxies: galaxy interactions trigger gas infall in the
  bar, leading to repetitive starbursts in the spiral galaxy
  disk, followed by subsequent fading. However, there exist also non-barred S0s having significant
  star formation in their inner rings, if which NGC 4138 is an example
  (Pogge $\&$ Eskridge 1993; see also Grouchy et al. 2010). It is
  possible that galaxy interactions/mergers play an important role
  also in the formation of these galaxies, in a sense that gas rich
  small companions might be swallowed by the more massive spiral
  galaxies. If the galaxy had an inner ring, the gas might have fallen
  into the potential well of the ring leading to a starburst in the
  ring. NGC 4138 has also a counter-rotating component (Jore et
  al. 1996; see also discussion in Comer\'on at al. 2010), which
  supports the merger hypothesis.  Or alternatively, star forming
  rings were formed in a process, where vertical satellite collision
  triggers the star forming ring (Mapelli et al. 2008).
 
%3. The sequence NGC 6654 -$>$ NGC 1533 -$>$ NGC 2983 (Fig. 13c) is an example of
%a {\it possible bar induced outer lens}. The R' type ring in NGC 6654
%is most probably a resonance ring, triggered by the main bar. If there
%is no mechanism maintaining the ring in the bar resonancei (***meaning?***),
%when evolved over time the ring becomes more dispersed (NGC 1533),
%finally having the appearance of an outer lens (NGC 2983).
%
%****Thus, a highly evolved ring. The L in NGC 2983 looks more like an evolved version
%of that seen in NGC 1326. Also, the R' feature in NGC 6654 is not typical of what is
%often seen in Sa galaxies. ***

\subsection{Prototypical multiple bar/lens structures}

The galaxies NGC 1543, NGC 6782 and NGC 3081 are {\it prototypical
  examples of double barred galaxies} (Fig. 15a). The scales in the
upper panels are selected to illustrate the main bars and lenses,
whereas the lower panels show the nuclear bars and lenses in the same
galaxies. NGC 1543 has two bars and lenses surrounding the bars,
extending to the same radius as the bar.  Intuitively it seems
plausible that the lenses were triggered by the bars. NGC 6782 also
has two bars, but a ring/lens is surrounding the main bar, and a
nuclear ring surrounds the nuclear bar. It is possible that the galaxy
also has a weak lens inside the nuclear ring, but that is difficult to
verify.  NGC 3081 has two weak bars, but in this galaxy the nuclear
and inner rings at the outskirts of the two bars are the dominant
features.  NGC 1317 is also double-barred. We find two nearly
orthogonal nuclear rings with a nuclear bar and nuclear lens inside
these features.

 In Figure 15b, {\it prototypical examples of multiple lenses in
   non-barred galaxies} are shown. The lenses can be intermediate
 types between rings and lenses as in NGC 3032, or full lenses as in
 NGC 524 and NGC 7192.  As we previously noted, NGC 524 is an example
 of a non-barred S0 galaxy having a series of circular lenses, the
 three lenses being clearly visible in this case. Depending on the
 prominence of the lenses the galaxies are classified either S0$^o$ or
 S0$^-$. These kind of galaxy illustrates interesting borderline cases
 between S0s and ellipticals: if the lenses are weak the galaxies can
 be easily misclassified as elliptical galaxies, because their surface
 brightness profiles are fairly similar.  For a full discussion of
 their nature, kinematic observations are also needed.

There are many questions related to multiple lenses in S0s that need
to be answered. For example:   (1) are lenses primarily formed soon after the disc formation or are they
rather bar-related products of secular evolution in galaxies.
(2) If produced mainly by secular evolution, are lenses 
former bars dissolved into lenses, or more likely structures triggered 
by bars, for example via ring formation?  
(3) What are the possible secular evolutionary
processes producing the multiple lenses in non-barred galaxies?  And
finally, (4) How can the multiple bar/lens systems be maintained in the
current hierarchical picture of galaxy formation? We discuss these 
issues in a forthcoming paper.

%Can multiple lenses in non-barred galaxies be formed by bars?
%That would be possible if bars in galaxies like NGC 1533 (see Fig. 11c)
%would slowly die away. In that case 'barlens' in NGC 1533 would make
%a lens, and 'RL' an outer lens, similar to those detected for example
%in NGC 524 and NGC 7192. 

\subsection{S0c Galaxies?}

Candidates of S0$_c$ type galaxies, which group of galaxies was
suggested by van den Bergh (1976), were searched from the NIRS0S
sample. The appearance of this kind of galaxies having disks with no
spiral arms, but bulge-to-total ($B/T$) flux ratios as small as
typically found in Sc type spirals, was noticed by Erwin et
  al. (2003) and Laurikainen et al.  (2006, 2010). In Figure 16 we
show representative examples of these galaxies.  Using the structure
decompositions of Laurikainen et al. (2010), and allowing for $B/T$
$\leq$ 0.1, 14 S0s were found.  This limit was selected because it is
the mean $B/T$ value for Sc type spirals, based on the decompositions
for spirals, made in a similar manner as for the NIRS0S galaxies. The
bulge flux is taken to be that fitted by a Sers\'ic function, whereas
the disk flux is a sum of all the disk components, including bars and
lenses. All the galaxies in Fig. 16 have a small bulge manifested as a
narrow peak in the surface brightness profile, and a prominent
extended disk. These galaxies have similar, or at most only slightly
fainter total absolute $K$-band magnitudes (for the 9 galaxies in the
figure $<M_K>$ = $-$23.6 mag, whereas for the 14 galaxies $<M_K>$
= $-$23.9) than the S0 galaxies in general ($<M_K>$ $\sim$
$-$24.0). The absolute magnitudes were calculated using the $K$-band
magnitudes from 2MASS, corrected for Galactic extinction taken from
NED, based on the maps of Schlegel, Finkbeiner $\&$ Davies (1998), and
using galaxy distances from the Catalog of Nearby Galaxies by Tully
(1988).  A Hubble constant of $H_o$=75 km s$^{-1}$ Mpc$^{-1}$ is used.

Of the non-barred galaxies (see Fig. 16) NGC 4138 and NGC 5273 have very
narrow peaks in the surface brightness profiles, and either ring or
subtle spiral features, which might be manifestations of earlier
spiral stage of these galaxies. NGC 1411 has more mass in the central
regions, due to a very prominent lens. The three barred galaxies, NGC
3081, NGC 4429 and NGC 4220 have obviously very small bulges embedded
in large disks. The last three galaxies, NGC 2983, NGC 3892 and NGC
5838, are examples of barred galaxies having prominent barlenses, the
small bulges embedded inside these lenses. Our example galaxies can be
former Sc spirals in which gas is either stripped or consumed by star
formation, as originally suggested by Baade (1963) and van den Bergh (1976):
due to dynamical heating of the disk the spiral arms have disappeared, but
the other disk structures like bars, lenses and rings are still visible, 
in a similar manner as in spirals.

\section{CONCLUSIONS} 

The NIRS0S atlas of 206 early-type disk galaxies is presented in the
K$_{\rm s}$-band, including 160 S0-S0/a galaxies. In order to discuss
the borderline of S0s with ellipticals and spirals, late-type
ellipticals classified as S0s in the RSA, and Sa spirals were also
included in the sample. A sub-sample of 185 galaxies forms a
magnitude-limited sample, having total magnitudes of $B_{\rm T}$
$\leq$ 12.5 mag and inclinations less than 65$^o$. The obtained images
are deep, typically reaching a surface brightness level of 23.5 mag
arcsec$^{-2}$ (exceptions are galaxies having too small FOV).  A
sub-arcsecond pixel scale ($\sim$ 0.25'') was used and the
observations were generally carried out in good seeing conditions
(FWHM $\sim$ 1''). The flux calibrated images are shown in many
different scales, optimized to show the multi-component nature of many
of the galaxies. In the Atlas panels the radial profiles of the
position angle, the ellipticity, and the deviation of the isophotes
from perfect ellipticities (b4) are also shown.

A detailed morphological classification was made using the criteria of
de Vaucouleurs (1959).
%but extending that work as described in the dVA and in Buta (2011). 
Special attention was paid to the recognition of
lenses in NIRS0S galaxies, which has been done in more systematic
manner than in any of the previous studies. Lenses are coded
in a similar manner as nuclear, inner and outer rings were previously
coded by Buta et al. (2010). A new lens type called a
'barlens', was also introduced, referring to the intermediate-sized,
bulge-looking component that seems prominent in many early-type barred
galaxies, presumably forming part of the bar itself. When elongated
along the bar it has the appearance of the so called 'boxy bar'.
Bar morphology is included in the classification: 
ansae morphology was detected in 33 bars, and x-shaped bar structure in 9
non-edge-on galaxies.

Besides visual classifications we
present also photometric classifications, which means that exponential
outer disks or faint inner structures were considered even if they
were not directly visible in the images.  The faint structures were
identified from unsharp masked or residual images after subtracting a
bulge model taken from the structural decompositions of Laurikainen et
al. (2010). Our visual and photometric classifications deviate for 42
galaxies: for example, 15 faint bars, outshone by bulges in visual
classification, were detected, and 7 elliptical galaxies were moved
into the S0 stage. However, the mean Hubble stage in our visual
classification in the near-IR is the same as that in the RC3,
determined in the optical wavelength range.

We confirm the previous result by \citet{lauri2009} that most
early-type disk galaxies have lenses, which we find to be the case in
both barred (61$\%$) and non-barred (38$\%$) galaxies. Most
importantly, we find that up to 25$\%$ of the Atlas galaxies,
including the S0-S0/a galaxies, have multiple lenses. However, only
six galaxies (4$\%$) have shells or ripples, which are expected to be
direct manifestations of recent mergers. The detection of multiple
lenses in a large number of S0-S0/a galaxies is a challenge to the
hierarchical formative processes of galaxies: it needs to be explained
how such lens systems were formed and survived in the merger events
that galaxies might have suffered several times in their lifetimes. We
discuss tentative morphological sequences of possible formative
processes of lenses. Possible candidates of S0$_c$ galaxies are shown,
which galaxies are expected to be former Sc-type spirals stripped out
of gas.
%However, in our photometric
%classification the Hubble stage was changed towards a later stage for
%15 galaxies, including 7 elliptical galaxies which were classified as
%S0s by us. Also, in 14 galaxies faint bars were detected in galaxies
%for which the bars were not visually seen.

\section*{Acknowledgments}

We acknowledge of significant observing time allocated to this project
during 2003-2009, based on observations made with several telescopes.
They include the New Technology Telescope (NTT), operated at the
Southern European Observatory (ESO) at La Silla in Chile. ESO is
supported by 15 countries: Austria, Belgium, Brazil, the Czech
Republic, Denmark, France, Finland, Germany, Italy, the Netherlands,
Portugal, Spain, Sweden, Switzerland and the United Kingdom.  Included
are also the William Herschel Telescope (WHT), the Italian Telescopio
Nazionale Galileo (TNG), and the Nordic Optical Telescope (NOT),
operated on the island of La Palma by, respectively, the Isaac Newton
Group of Telescopes, the Fundaci\'on Galileo Galilei of the INAF
(Istituto Nazionale di Astrofisica), and jointly by Denmark, Finland,
Iceland, Norway, and Sweden, in the Spanish Observatorio del Roque de
los Muchachos of the Instituto de Astrof\'\i sica de Canarias.
Included are also the 4m telescope operated at the Cerro Tololo
Inter-American Observatory (CTIO) in La Serena, Chile, and Flamingos
(FLMN) operated at KPNO (Kitt Peak National Observatory), in Tucson,
Arizona.  We also acknowledge Jarkko Laine, Sebastien Comer\'on, Sami
Airaksinen, Tom Speltincx, Leena Pelttari and Timothy Brockett, who have participated in
making the observations.  This publication makes use of data products
from the Two Micron All Sky Survey, which is a joint project of the
University of Massachusetts and the Infrared Processing and Analysis
Center/California Institute of Technology, funded by the National
Aeronautics and Space Administration and the National Science
Foundation. This research has made use of the NASA/ IPAC Infrared
Science Archive, which is operated by the Jet Propulsion Laboratory,
California Institute of Technology, under contract with the National
Aeronautics and Space Administration. We acknowledge Academy of
Finland of significant financial support.  RB acknowledges the support
of NSF grant AST 050-7140 for part of this work.

\clearpage
\newpage

\begin{table}
\begin{center}
\caption{The observing campaigns.  \label{tbl-1}}
\begin{tabular}{lllll}
\hline
 \noalign{\smallskip}         
  Tel.+instrument & date             & resolution  & FOV       \\% & gain        
                  &                  & [arcsec/pix]& [arcmin]   \\%& [e$^-$/ADU]             
\hline
 \noalign{\smallskip}
Near-IR:          &                  &             &           \\ %&          
  \noalign{\smallskip}   

NOT(2.5m)/NOTCam        & 17-20 Jan 2003   & 0.233       & 4.0 x 4.0  \\% & 3.1        
NOT(2.5m)/NOTCam        & 28-29 Sept 2003  & 0.233       & 4.0 x 4.0  \\% & 3.1     
NOT(2.5m)/NOTCam        & 08-11 Jan 2004   & 0.233       & 4.0 x 4.0  \\% & 3.1     
NTT(3.6m)/SOFI          & 20-23 Dec 2004   & 0.288       & 4.9 x 4.9  \\% & 5.4    
NOT(2.5m)/NOTCam        & 19-23 May 2005   & 0.233       & 4.0 x 4.0 \\%  & 3.1     
WHT(4.2m)/LIRIS         & 11-14 May 2006   & 0.250       & 4.3 x 4.3  \\% & 3.7    
NTT(3.6m)/SOFI          & 17-21 July 2006  & 0.288       & 4.9 x 4.9  \\% & 5.4    
WHT(4.2m)/LIRIS         & 03-05 March 2007 & 0.250       & 4.3 x 4.3  \\% & 3.7    
CTIO(4m)                & 27-29 Oct 2007   & 0.306       & 5.2 x 5.2  \\% & ?      
TNG(3.6m)/NICS          & 17-19 Nov 2007   & 0.252       & 4.2 x 4.2  \\% & 8.0      
%NOT(2.5m)/NOTCam        & 16-17 Jul 2007   & 0.233       & 4.0 x 4.0  \\% & 3.1    
TNG(3.6m)/NICS          & 16-17 Jun 2008   & 0.252       & 4.2 x 4.2  \\% & 8.0    
NTT(3.6m)/SOFI          & 10-14 Jun 2008   & 0.288       & 4.9 x 4.9  \\% & 5.4     
WHT(4.2m)/LIRIS         & 02-03 Dec 2008   & 0.250       & 4.3 x 4.3  \\% & 3.7    
%TNG(3.6m)/NICS          & 05-08 Jan 2009   & 0.252       & 4.2 x 4.2  \\% & 8.0    
TNG(3.6m)/NICS          & 10-12 Apr 2009   & 0.252       & 4.2 x 4.2  \\% & 8.0    
KPNO(2.1m)/FLMN         & 04-07 May 2009   & 0.606       & 19.5 x 19.5 \\%& 4.9    
                  &                  &             &             &     \\%  
\hline
\end{tabular}
\end{center}
\end{table}

\clearpage
\newpage

\begin{table}
\begin{center}
\caption{The image quality. Galaxy identification and telescope are
  indicated, together with Full-Width at Half-Maximum (FWHM) and
  estimated 1-$\sigma$ sky variation ($\sigma_{sky}$). $\Delta \mu_0$
  indicates the difference of the zeropoint derived for the galaxy
  based on 2MASS $m_{14}$, with respect to the campaign mean for the
  same airmass, $\Delta k_{20}$ and $\Delta k_{ext}$ indicate the differences
  of the measured isophotal magnitudes with respect to 2MASS values.  
\label{tbl-2}}
\begin{tabular}{llllrrrl}
\hline
 \noalign{\smallskip}         
Galaxy & telescope & FWHM       & $\sigma_{\rm sky}$     & $\Delta \mu_0$ & $\Delta k_{20}$ & $\Delta k_{\rm ext}$   &    \\
       &            &  [arcsec]  & [mag/arcsec$^2$]  & [mag]          & [mag]          & [mag]             &     \\
 (1)   &   (2)      &  (3)      &  (4)              &  (5)           &  (6)           & (7)              & \\     
\hline
 \noalign{\smallskip}
ESO 137-10   &   NTT &  1.25 &  22.7 &   0.06 &  -0.13 &  -0.15 &   \\
ESO 337-10   &   NTT &  0.89 &  22.2 &  -0.00 &  -0.02 &   0.00 &   \\
ESO 208-21   &   NTT &  0.63 &  22.2 &   0.11 &  -0.03 &  -0.04 &   \\
IC 1392      &   TNG &  0.97 &  22.9 &  -0.06 &  -0.04 &  -0.03 &   \\
IC 4214      &   NTT &  0.84 &  21.9 &  -0.02 &  -0.02 &   0.01 &   \\
\hline
\end{tabular}
\end{center}
\end{table}

\begin{table}
\begin{center}
%\caption{Continue  }
\begin{tabular}{llllrrrl}
\hline
 \noalign{\smallskip}         
 Galaxy & telescope & FWHM       & $\sigma_{\rm sky}$     & $\Delta \mu_0$ & $\Delta k_{20}$ & $\Delta k_{\rm ext}$   &    \\
      &            &  [arcsec]  & [mag/arcsec$^2$]  & [mag]          & [mag]          & [mag]             &     \\
(1)   & (2)        &  (3)       &  (4)              &  (5)           & (6)            & (7)              & \\
\hline
 \noalign{\smallskip}
NGC 2782     &   WHT &  1.12 &  22.7 &  -0.09 &   0.01 &   0.04 &   \\
NGC 2787     &   NOT &  1.10 &  21.3 &   0.05 &  -0.02 &  -0.02 &   \\
NGC 2855     &   NOT &  0.93 &  21.7 &   0.01 &  -0.03 &  -0.00 &   \\
NGC 2859     &   NOT &  0.93 &  21.7 &  -0.03 &  -0.03 &   0.13 &   \\
NGC 2880     &   WHT &  0.93 &  22.9 &  -0.06 &   0.03 &   0.07 &   \\
\hline
\end{tabular}
\end{center}
\end{table}

\begin{table}
\begin{center}
%\caption{Continue}
\begin{tabular}{llllrrrl}
\hline
 \noalign{\smallskip}         
 Galaxy & telescope & FWHM       & $\sigma_{\rm sky}$     & $\Delta \mu_0$ & $\Delta k_{20}$ & $\Delta k_{\rm ext}$   &    \\
      &            &  [arcsec]  & [mag/arcsec$^2$]  & [mag]          & [mag]          & [mag]             &     \\
(1)   &   (2)      &  (3)       & (4)               &  (5)           & (6)            & (7)               &  \\
\hline
 \noalign{\smallskip}
NGC 4459     &   WHT &  0.68 &  23.0 &  -0.01 &  -0.09 &  -0.09 &   \\
NGC 4472     &  KPNO &  2.30 &  21.2 &   0.04 &   0.15 &   0.32 &   \\
NGC 4474$^c$ &   TNG &  1.36 &  22.1 &        &        &        &   \\
NGC 4477     &   WHT &  1.00 &  23.5 &  -0.02 &   0.02 &   0.04 &   \\
NGC 4503     &   TNG &  1.26 &  22.6 &   0.06 &   0.01 &   0.00 &   \\
\hline
\end{tabular}
\end{center}
\end{table}

\begin{table}
\begin{center}
%\caption{Continue  }
\begin{tabular}{llllrrrl}
\hline
 \noalign{\smallskip}         
  Galaxy & telescope & FWHM       & $\sigma_{\rm sky}$     & $\Delta \mu_0$ & $\Delta k_{20}$ & $\Delta k_{\rm ext}$   &    \\
      &            &  [arcsec]  & [mag/arcsec$^2$]  & [mag]          & [mag]          & [mag]             &     \\
 (1)  &  (2)       &  (3)       & (4)               & (5)            & (6)            & (7)               &  \\
\hline
 \noalign{\smallskip}
NGC 6782     &   NTT &  0.63 &  22.3 &  -0.05 &  -0.05 &  -0.06 &   \\
NGC 6861     &   NTT &  0.66 &  22.4 &  -0.05 &  -0.03 &  -0.00 &   \\
NGC 6958     &  CTIO &  0.91 &  22.0 &  -0.03 &   0.01 &   0.06 &   \\
NGC 7029     &   NTT &  0.69 &  22.4 &  -0.07 &  -0.07 &  -0.05 &   \\
NGC 7049     &   NTT &  0.72 &  22.4 &   0.06 &  -0.01 &   0.07 &   \\
NGC 7079     &   NTT &  0.66 &  22.4 &  -0.07 &  -0.04 &  -0.09 &   \\
NGC 7098     &   NTT &  0.84 &  21.7 &   0.01 &  -0.21 &  -0.22 &   \\
NGC 7192     &   NTT &  0.81 &  22.1 &  -0.05 &   0.06 &   0.11 &   \\
NGC 7213     &   NTT &  0.75 &  22.3 &  -0.03 &  -0.05 &  -0.06 &   \\
NGC 7217     &  KPNO &  2.42 &  21.2 &  -0.01 &   0.13 &   0.18 &   \\
NGC 7332     &   WHT &  1.00 &  22.9 &  -0.02 &  -0.02 &  -0.03 &   \\
NGC 7339     &   WHT &  0.95 &  22.1 &  -0.00 &  -0.04 &  -0.05 &   \\
NGC 7371     &   NTT &  0.60 &  22.3 &  -0.05 &   0.04 &   0.06 &   \\
NGC 7377     &   NOT &  0.91 &  21.0 &  -0.04 &   0.00 &   0.02 &   \\
NGC 7457     &   WHT &  0.93 &  22.6 &  -0.01 &  -0.06 &  -0.07 &   \\
NGC 7585     &   NTT &  1.11 &  22.7 &   0.05 &   0.07 &   0.15 &   \\
NGC 7727     &   NTT &  0.90 &  22.8 &   0.01 &  -0.01 &   0.06 &   \\
NGC 7742     &   NOT &  0.77 &  21.1 &   0.03 &  -0.01 &   0.01 &   \\
NGC 7743     &   NOT &  0.82 &  20.9 &   0.24 &   0.09 &   0.08 &   \\
NGC 7796     &   NTT &  0.72 &  22.3 &  -0.05 &  -0.01 &   0.02 &   \\
\hline
\end{tabular}
\end{center}
$^a$ 2MASS data missing - use campaign zeropoint

$^b$ interacting -use campaign zeropoint

$^c$ 2MASS data inconsistent - use campaign zeropoint
\end{table}

%d AAT did not check campaign

\clearpage
\newpage

%new component type added: 'bl'= lens inside the bar, which lens is larger than nuclear lens.
% new component type added: 'ob' = outer bar, which is the largest bar of the three.
\begin{table}
\begin{center}
\caption{Visual classification in the near-IR, compared with optical classification (RC3 and RSA). \label{tbl-3}}
\begin{tabular}{llllcc}
\hline
 \noalign{\smallskip}         
%         &               &             &                           &         &        &            \\
%\tableline\tableline
Galaxy       & type (2.2$\mu$m)  &type (RC3) & type (RSA) & T (2.2$\mu$m) & T (RC3) \\ % & Comment        \\% & Tel  & FWHM     \\  %& Filters       \\
(1)          & (2)               & (3)       & (4)        & (5)            & (6)     \\ %  & (7)     \\% & (5)     & (6)     \\  %& (8)     &        \\
%\tableline
\hline
 \noalign{\smallskip}

      IC 1392 *   & E(b)4/S0$^-$ sp                                               & .L..-*. &             & $-$4.0  & $-$3.0  \\ % &    boxy (b4-max)        \\ %  & TNG     & 0.97      %& K$_s$   \\
      IC 4214 * $^1$ & (R$_1$)S$\underline{\rm A}$B($\underline{\rm r}$l,nr)0$^+$ & PSBR2.. &            &  $-$1.0 &  1.5  \\ %   & better scale: nb           \\ %& NTT     & 0.84  \\     %& K$_s$  \\
      IC 4329 *    & SA0$^o$ shells/ripples                                     & .LXS-.. & S01(5)     & $-$2.0  & $-$3.0  \\ %  & oval bar:   ?   \\ %    & NTT     & 0.78  \\    %&  K$_s$,B,V  \\ 
      IC 4889 *    & SA0$^-$                                                     & .E.5+.. & S01/2(5)   &  $-$5.0 & $-$5.0  \\ %  & decomp: expo disk -$>$ S0 \\%   & CTIO    & 0.83  \\    %& K$_s$,B,V    \\  
      IC 4991 * $^1$   & coreE                                                   & .LAR0?P &             & $-$5.0  & $-$2.0   \\ %  & decomp: expo disk -$>$ S0\\%   & NTT     & 0.75  \\    %& K$_s$,B,V      \\
      IC 5240 *    & SB$_x$(r)a                                                 & .SBR1.. & SBa(r)      &   1.0   &  1.0   \\
\hline
\end{tabular}
\end{center}
\end{table}

\newpage
\clearpage

\begin{table}
\begin{center}
%\caption{Continue}
\begin{tabular}{llllccl}
\hline
 \noalign{\smallskip}         
Galaxy       & type(2.2$\mu$m) & type(RC3)& type(RSA)   & $T$(2.2$\mu$m) & $T$(RC3)\\%  & comment         \\% & Tel  & FWHM     \\  %& Filters       \\
(1)          & (2)             & (3)      & (4)         & (5)            & (6)     \\%  & (7)    \\% & (5)     & (6)     \\  %& (8)     &        \\
\hline
 \noalign{\smallskip} 

      NGC 2273 *   & (R)SAB(rs,bl,nb)a                                          & .SBR1*. &                   &   1.0   &   .5     \\%  & (Ron's class:  (R)SAB(rs,nlb)a)        \\%  & NOT    & 1.32  \\    % &  K$_s$^1     \\ 
      NGC 2292 * $^1$   & E: pec                                                & .LX.0P. &                     & $-$5.0  & $-$2.0  \\%   & decomp: expo disk -$>$ S0  \\%     & TNG  IA  & 1.39 \\     % &  K$_s$    \\
     NGC 2293 *   & SAB$_a$(bl)0/a                                                & .LXS+P. &                    &  0.0    & $-$1.0  \\%   & \\%  & TNG    & 1.39 \\    % & K$_s$  \\ 
     NGC 2300 *   & (R$^{\prime}$L)SA(sl)0$^o$                 & .LA.0.. &  E3                &  $-$2.0 & $-$2.0   \\%  &      \\%  & TNG    & 1.91 \\    % & K$_s$   \\
     NGC 2380 *   & SA(l,nl)0$^-$                                                 & .LX.0*. &                    & $-$3.0  & $-$1.7   \\%  & $^2$PA/ell-profile: multiple lenses    \\%  & TNG    & 1.31 \\     % &  K$_s$ \\
      NGC 2460     & S$\underline{\rm A}$B(rs)a                                 & .SAS1.. &  Sab(s)            &   1.0   &  1.0    \\
\hline
\end{tabular}
\end{center}
\end{table}

\newpage
\clearpage

\begin{table}
\begin{center}
%\caption{Continue}
\begin{tabular}{llllccl}
\hline
 \noalign{\smallskip}         
Galaxy       & type(2.2$\mu$m)  & type(RC3) &type(RSA)   & $T$(2.2$\mu$m) & $T$(RC3)\\%  & Comment      \\% & Tel  & FWHM     \\  %& Filters       \\
(1)          & (2)              & (3)        & (4)       & (5)            & (6)     \\%  & (7)       \\% & (5)     & (6)     \\  %& (8)     &        \\
\hline
 \noalign{\smallskip}

     NGC 4106 *  & SAB(s,nl)a                                                & .LBS+.. & SB0/a (tides)        & 1.0    & $-$1.0 \\%  &  nb     \\%  & NTT     & 1.12 \\   %& K$_s$ \\ 
   NGC 4138 *  & SA(r)0$^+$                                                  & .LAR+.. & Sab(r)              & $-$1.0  & $-$1.0 \\%  &  \\ %lens   \\%  & NOT     & 1.04 \\    % & B,V,K$_s$   \\
    NGC 4143 *  & (L,R$^{\prime}$L)SAB$_a$(s,nb,bl)0$^-$                                        & .LXS0.. &  S0$_1$(5)/Sa       & $-$3.0  & $-$2.0  \\%  &   \\%  & TNG     & 1.16 \\    %& B,V,K$_s$   \\
     NGC 4150 * $^1$  & (L)SA(nl)0$^-$                                              & .LAR0\$.&  S0$_3$(4)/Sa        & $-$3.0  & $-$2.0 \\%  &  unsharp mask: nr  \\%  & WHT     & 1.12 \\   % & K$_s$      \\
     NGC 4203 *  & (L,R$^{\prime}$L)S$\underline{\rm A}$B$_a$(l,rl)0$^-$               & .LX.-*. & S0$_2$(1)           & $-$3.0  & $-$3.0 \\%  &  residual: bar is clear \\%  & WHT     & 1.32 \\     %& K$_s$      \\ 
    NGC 4220    & (RL)SAB$_x$(r)0$^+$ sp                                     & .LAR+.. & Sa(r)  S R1         & $-$1.0  & $-$1.0 \\
\hline
\end{tabular}
\end{center}
\end{table}

\clearpage
\newpage

\begin{table} 
\begin{center}
%\caption{Continue}
\begin{tabular}{llllccl}
\hline
 \noalign{\smallskip}         
Galaxy       & type(2.2$\mu$m)   & type(RC3) &  type(RSA)  & $T$(2.2$\mu$m) & $T$(RC3) \\% & Comment        \\% & Tel  & FWHM     \\  %& Filters       \\
(1)         & (2)              & (3)      & (4)            & (5)            & (6)      \\% & (7)         \\% & (5)       & (6)     \\  %& (8)     &        \\
\hline
 \noalign{\smallskip} 

     NGC 5026 *  & (L)SB(rs,nl,bl)a                                         & PSBT0.. &                       &  1.0     &  $-$.2 \\%   &   \\% & NTT     & 0.82 \\  %& K$_s$,B,V  \\
     NGC 5078 *  & Sa sp                                                 & .SAS1*/ &                       &   1.0    &  1.0   \\%  &         \\%  & KPNO    & 1.81 \\    % & K$_s$ \\
    NGC 5087 * $^1$   &  E2/SA0$^-$ sp                                    & .LA.*.. & S0$_3$(5)             & $-$4.0   & $-$3.0 \\%  &     \\% & NTT     & 1.06 \\    %& K$_s$  \\
     NGC 5101 *  & (RR)SB($\underline{\rm r}$s,nl,bl)0/a                 & RSBT0.. & SBa                   &   0.0   &   .0   \\% &         \\%  & use AAT & -    \\    % &     \\
     NGC 5121 *  & (RL)SA(rl)0$^+$                                       & PSAS1.. & S0$_1$(4)/Sa          &$-$1.0    &  1.0    \\% &  \\%  & NTT     & 0.86 \\   % & K$_s$  \\
     NGC 5206 *  & dSB0,N                                                & PLBT0P* &                       & $-$2.0   & $-$2.5  \\
\hline

\end{tabular}
\end{center}
\end{table}

\clearpage
\newpage
\begin{table}
\begin{center}
%\caption{Continue}
\begin{tabular}{llllccl}

\hline
 \noalign{\smallskip}         
Galaxy       & type(2.2$\mu$m)   & type(RC3)  & type(RSA)  & $T$(2.2$\mu$m) & $T$(RC3)\\%  & Comment        \\% & Tel  & FWHM     \\  %& Filters       \\
(1)         & (2)              & (3)      & (4)            & (5)            & (6)     \\%  & (7)       \\% & (5)     & (6)     \\  %& (8)     &        \\
\hline
 \noalign{\smallskip}

%      NGC 7184   & SB(r)b                   &  -      &  -    \\ %& B,V    \\
      NGC 7192 *  & (L)SA(l)0$^-$                                         & .E+..*. &  S0$_2$(0)              & $-$3.0   & $-$4.3  \\%  & $^7$decom: lenses manifested as ... \\%as expo subsection in the m$\mu$ profile\\%              & NTT     & 0.81  \\  % & K$_s$,B,V        \\
      NGC 7213 * $^1$   & SA(r,rl)0$^o$                                   & .SAS1*. & Sa(rs)                & $-$2.0   &  1.0   \\%  &  $^7$decomp: lenses manifested as ... \\% as expo subsections in the m$\mu$ profile \\%   & NTT    & 0.75  \\   % & K$_s$        \\
     NGC 7217    & (R$^{\prime}$)SA(l,nl)0/a                               & RSAR2.. & Sb(r)II-III             &  0.0    &  2.0    \\%  &    \\%     & KPNO   & 2.40  \\   % &  K$_s$ \\
%    NGC 7252     & merger M           & - ??                             & -      & &  -    \\  %& B,V     \\
      NGC 7332    & SB$_x$0$^o$ sp                                        & .L...P/ & S0$_{2/3}$(8)              & $-$2.0  & $-$2.0 \\%  &  \\%    & WHT     & 1.00  \\  %& K$_s$      \\
      NGC 7339    & SA(s)bc sp                                            & .SXS4*\$ &                         &  4.0    &  4.0    \\%  &           \\%    & WHT     & 0.96  \\  %& K$_s$      \\ 
     NGC 7371 *  &  SA$\underline{\rm B}$(s)a                             & RSAR0*. & SBa(r)II                 &  1.0    &   .0   \\
\hline

\end{tabular}
\end{center}
%\end{table}
%\clearpage

%$^1$ $^2$ These galaxies have published $K_s$-band observations in Laurikainen et al. 2005, 2006, respectively. 

* Form part of the complete magnitude-limited NIRS0S sample.

%$^3$ The FOV of our image is not ehough to confirm whether the galaxy has as exponential
%     disk or not.
%$^3$ Some signatures of outer spirals (see Fig. X)

%Otokseen kuuluvat lisaksi nama:
%(*)     eso137-34 - probably a distant galaxy with bright stars in the field (.SXS0?.)
%(*)      ic5250/50A - merger (.L...P?/.L...P?)
%(*)      ngc147   -dE (.E.5.P.)
%(*)      ngc185   - dE (.E.3.P.)
%(*)      ngc205  - could have been observed making a mosaic (.E.5.P.)
%(*)      ngc404  - bright star makes impossible to observe the galaxy (.LAS-*.)
%*      ngc1291 - but we have SINGS image (RSBS0..)
%*      ngc1316 - but we have SINGS image (.LXS0P.)
%*      ngc1546 - simply not observed (.LA.+?.)	
%*      ngc1808 - could have been observed making a mosaic (RSXS1..)
%*      ngc1947 - simply not observed (.L..-P.)
%*      ngc5128  - large nearby early-type with dust lanes (.L...P.)
% 
%
%ESO 208$-$G21\tablenotemark{1}  & E(d)5                                                              & SAB$^-$                  & $K_s$  & 0.63    &1.1 \\
%NGC 1079                        & (R$_1$R$_2^{\prime}$)S$\underline{\rm A}$B($\underline{\rm r}$s)a  &  (R)SAB(rs)0/a        & $K_s$  & 0.84    &1.0  \\
%Revised Shapley-Ames Catalog of Bright Galaxies (Sandage, 1981) 

\end{table}

\clearpage
\newpage

% Photometric classification includes: mu-profiles, ell-fitting, images in different scales, (evidence also from decompositions)
\begin{table}
\begin{center}
\caption{Photometric classification, if not the same as visual classification in Table 3. \label{tbl-4}}
\begin{tabular}{lll}
\hline
 \noalign{\smallskip}         
%\tableline\tableline
Galaxy       & type (2.2$\mu$m) &   $T$(2.2$\mu$m)   \\
(1)          & (2)              & (3)       \\
%\tableline
\hline
 \noalign{\smallskip} 

   IC 4214    & (R$_1$)S$\underline{\rm A}$B($\underline{\rm r}$l,nr,nb)0$^+$  &  -1     \\
%   IC 4889    & SA0$^-$                                                         & -3    \\
   IC 4991    & SA0$^-$                                                         & -3   \\
   IC 5267    & (RL)SA(l,nr)0/a                                              & 0       \\
%   IC 5328    & SB(l)0$^-$     S$\underline{\rm A}$B(l)0$^-$                                              & -3      \\ 
   ESO 337-10 & SA0$^-$                                               &  -3       \\
   NGC 0484   & SA(nb)0$^-$                                              & -3    \\
   NGC 0507   & (L)SAB(nb)0$^-$                                               & -3    \\
\hline

\end{tabular}
\end{center}

\end{table}
\clearpage
\newpage

% length for N3619, N7727 cannot be measured.
% Sauvojen pituudet toistaiseksi pikseleina!
\begin{table}
\begin{center}
\caption{Ring and lens dimension. \label{tbl-5}}
\begin{tabular}{llccc}
\hline
 \noalign{\smallskip}         
%\tableline\tableline
Galaxy       & feature type & major axis radius &   q & PA \\
             &              & [arcsec]          &      & [degrees] \\
(1)          &  (2)         & (3)               & (4)   & (5)  \\
%\tableline
\hline
 \noalign{\smallskip}

IC 4214  & R$_1$                &     59.3$\pm$0.04 &  0.590$\pm$.001 &    179.8$\pm$ 0.06\\
IC 4214  & $\underline{\rm r}$l &     28.1$\pm$0.04 &  0.483$\pm$.001 &    159.9$\pm$ 0.05\\
IC 4214  & nr                   &      6.4$\pm$0.01 &  0.674$\pm$.001 &    156.6$\pm$ 0.15\\
%IC 4214  & nb                   &      4.8$\pm$0.03 &  0.254$\pm$.003 &     90.2$\pm$ 0.12\\
IC 5240  & r                    &     38.9$\pm$0.03 &  0.651$\pm$.001 &    107.0$\pm$ 0.06\\
IC 5267  & l                    &     44.7$\pm$0.10 &  0.740$\pm$.002 &    142.3$\pm$ 0.20\\
\hline
\end{tabular}
\end{center}
\end{table}

\clearpage
\newpage

\begin{table}
\begin{center}
%\caption{Ring, Lens, and Nuclear Bar Dimension. \label{tbl-4}}
\begin{tabular}{llccc}
\hline
 \noalign{\smallskip}         
%\tableline\tableline
Galaxy       & feature type & major axis radius & q & PA \\
             &              & [arcsec]          &                        & [degrees] \\
(1)          &  (2)         & (3)               & (4)   & (5)  \\
%\tableline
\hline
 \noalign{\smallskip} 
NGC 1440 & L                    &     43.7$\pm$0.03 &  0.753$\pm$.001 &     19.9$\pm$ 0.10\\
NGC 1440 & $\underline{\rm r}$s                   &     24.7$\pm$0.07 &  0.765$\pm$.002 &     33.8$\pm$ 0.32\\
NGC 1440 & bl                   &     10.0$\pm$0.02 &  0.804$\pm$.002 &     48.4$\pm$ 0.29\\
NGC 1452 & r                    &     48.7$\pm$0.04 &  0.596$\pm$.001 &    109.7$\pm$ 0.05\\
NGC 1452 & bl                   &     18.3$\pm$0.01 &  0.887$\pm$.001 &    103.2$\pm$ 0.32\\
\hline
\end{tabular}
\end{center}
\end{table}

\clearpage
\newpage

\begin{table}
\begin{center}
%\caption{Ring, Lens, and Nuclear Bar Dimension. \label{tbl-4}}
\begin{tabular}{llccc}
\hline
 \noalign{\smallskip}         
%\tableline\tableline
Galaxy       & feature type & major axis radius & q & PA \\
             &              & [arcsec]          &                        & [degrees] \\
(1)          &  (2)         & (3)               & (4)   & (5)  \\
%\tableline
\hline
 \noalign{\smallskip} 

NGC 2782 & nr                   &      5.6$\pm$0.01 &  0.464$\pm$.001 &     87.7$\pm$ 0.11\\
NGC 2787 & nrl                    &     45.5$\pm$0.08 &  0.531$\pm$.001 &    104.8$\pm$ 0.09\\
NGC 2787 & bl                   &     21.3$\pm$0.06 &  0.507$\pm$.002 &    106.3$\pm$ 0.14\\
NGC 2859 & R                    &    107.9$\pm$0.13 &  0.701$\pm$.001 &     83.2$\pm$ 0.15\\
NGC 2859 & r$\underline{\rm l}$ &     39.3$\pm$0.03 &  0.875$\pm$.001 &     84.8$\pm$ 0.28\\
\hline
\end{tabular}
\end{center}
\end{table}

\clearpage
\newpage

\begin{table}
\begin{center}
%\caption{Ring, Lens, and Nuclear Bar Dimension. \label{tbl-4}}
\begin{tabular}{llccc}
\hline
 \noalign{\smallskip}         
%\tableline\tableline
Galaxy       & feature type & major axis radius & q & PA \\
             &              & [arcsec]          &                        & [degrees] \\
(1)          &  (2)         & (3)               & (4)   & (5)  \\
%\tableline
\hline
 \noalign{\smallskip} 

NGC 3718 & rs                   &     88.4$\pm$0.11 &  0.736$\pm$.001 &    156.0$\pm$ 0.14\\
NGC 3718 & l                    &     60.8$\pm$0.07 &  0.823$\pm$.001 &    168.2$\pm$ 0.27\\
NGC 3718 & nl                   &     14.6$\pm$0.02 &  0.973$\pm$.002 &    119.1$\pm$ 2.14\\
NGC 3729 & r                    &     38.3$\pm$0.13 &  0.462$\pm$.002 &    159.8$\pm$ 0.12\\
NGC 3892 & L                    &     83.1$\pm$0.04 &  0.922$\pm$.001 &     27.8$\pm$ 0.23\\
\hline
\end{tabular}
\end{center}
\end{table}

\clearpage
\newpage

\begin{table}
\begin{center}
%\caption{Ring, Lens, and Nuclear Bar Dimension. \label{tbl-4}}
\begin{tabular}{llccc}
\hline
 \noalign{\smallskip}         
%\tableline\tableline
Galaxy       & feature type & major axis radius & q & PA \\
             &              & [arcsec]          &                        & [degrees] \\
(1)          &  (2)         & (3)               & (4)   & (5)  \\
%\tableline
\hline
 \noalign{\smallskip}

NGC 4340 & bl                   &     25.6$\pm$0.07 &  0.579$\pm$.002 &    104.4$\pm$ 0.17\\
NGC 4340 & nr                  &      7.3$\pm$0.04 &  0.662$\pm$.004 &    110.7$\pm$ 0.39\\
%NGC 4340 & nb                   &      5.1$\pm$0.17 &  0.452$\pm$.016 &     29.1$\pm$ 0.79\\
NGC 4369 & rs                   &      9.5$\pm$0.02 &  0.908$\pm$.003 &     65.7$\pm$ 0.87\\
NGC 4371 & r                    &     57.1$\pm$0.05 &  0.553$\pm$.001 &     91.2$\pm$ 0.07\\
NGC 4371 & nr                   &     10.9$\pm$0.03 &  0.474$\pm$.002 &     89.1$\pm$ 0.16\\
\hline
\end{tabular}
\end{center}
\end{table}

\clearpage
\newpage

\begin{table}
\begin{center}
%\caption{Ring, Lens, and Nuclear Bar Dimension. \label{tbl-4}}
\begin{tabular}{llccc}
\hline
 \noalign{\smallskip}         
%\tableline\tableline
Galaxy       & feature type & major axis radius & q & PA \\
             &              & [arcsec]          &                        & [degrees] \\
(1)          &  (2)         & (3)               & (4)   & (5)  \\
%\tableline
\hline
 \noalign{\smallskip}

NGC 5101 & OOR                  &    160.1$\pm$0.07 &  0.886$\pm$.000 &    148.7$\pm$ 0.12\\
NGC 5101 & IOR                  &    104.1$\pm$0.10 &  0.805$\pm$.001 &    166.0$\pm$ 0.18\\
NGC 5101 & $\underline{\rm r}$s &     52.1$\pm$0.03 &  0.823$\pm$.001 &    115.6$\pm$ 0.10\\
NGC 5101 & nl                   &      5.1$\pm$0.01 &  0.972$\pm$.004 &     42.0$\pm$ 3.65\\
NGC 5101 & bl                   &     30.5$\pm$0.07 &  0.975$\pm$.003 &     86.1$\pm$ 3.64\\
\hline
\end{tabular}
\end{center}
\end{table}

\clearpage
\newpage

\begin{table}
\begin{center}
%\caption{Ring, Lens, and Nuclear Bar Dimension. \label{tbl-4}}
\begin{tabular}{llccc}
\hline
 \noalign{\smallskip}         
%\tableline\tableline
Galaxy       & feature type & major axis radius & q & PA \\
             &              & [arcsec]          &                        & [degrees] \\
(1)          &  (2)         & (3)               & (4)   & (5)  \\
%\tableline
\hline
 \noalign{\smallskip} 

NGC 6012 & x1r                  &     21.9$\pm$0.50 &  0.254$\pm$.006 &    151.4$\pm$ 0.16\\
NGC 6340 & nl                   &      4.9$\pm$0.01 &  0.924$\pm$.001 &    101.1$\pm$ 0.63\\
NGC 6340 & l                    &     19.0$\pm$0.04 &  0.940$\pm$.003 &    105.0$\pm$ 1.61\\
NGC 6340 & RL                   &     32.2$\pm$0.05 &  0.887$\pm$.002 &    107.2$\pm$ 0.49\\
NGC 6438 & l                    &     15.7$\pm$0.05 &  0.927$\pm$.004 &    175.2$\pm$ 1.79\\
\hline
\end{tabular}
\end{center}
\end{table}

\clearpage
\newpage

% length for N3619, N7727 cannot be measured.
% Sauvojen pituudet toistaiseksi pikseleina!
\begin{table}
\begin{center}
\caption{Bar dimensions: major axis radius ($r_{\rm vis}$, $r_{\rm ell}$ $r_{\rm L}$, explained in Section 6), orientations (PA) and minor-to-major axis ratios (q) of bars. STdev
is the estimated standard deviation of $r_{\rm vis}$ and $r_{\rm ell}$. \label{tbl-6}}
\begin{tabular}{llllllll}
\hline
 \noalign{\smallskip}         
%\tableline\tableline
Galaxy  & bartype & PA       & $r_{\rm vis}$ & $r_{\rm ell}$ & $r_{\rm L}$ & STdev  & q \\% ell=1-b/a -> b/a=1-ell   
        &         &[degrees] & [arcsec]        & [arcsec]        & [arcsec]     & [arcsec]         &  \\
(1)     & (2)     & (3)      & (4)             & (5)             & (6)          &  (7)     & (8)      \\
\hline
 \noalign{\smallskip} 

IC 4214  & nb     &  89.3    &  4.1          &  3.9       &  4.3  &0.1   &0.750  \\%  .288 
         & $\underline{\rm A}$B     & 161.7    & 30.8          & 26.2       & 30.5  &2.3   &0.468\\
% \noalign{\smallskip} 
IC 5240  & B$_x$  &  93.4    & 34.1          & 38.6       &        &2.2   &0.269 \\ %   .288 
% \noalign{\smallskip} 
IC 5328  & $\underline{\rm A}$B$_a$      &  41.0    & 50.4          & 42.5       & 51.2  & 3.0  &0.495\\%  .288
% \noalign{\smallskip} 
NGC 474  & AB     &  28.6    & 18.0          & 18.0       &        & 0.0  &0.750\\
\hline
\end{tabular}
\end{center}
\end{table}

\begin{table}
\begin{center}
%\caption{Continue  }
\begin{tabular}{llllllll}
\hline
 \noalign{\smallskip}         
Galaxy   & bartype & PA      &r$_{\rm vis}$       & r$_{\rm ell}$     & r$_{\rm L}$  & STdev  & q \\% ell=1-b/a -> b/a=1-ell     
         &         &[degrees]& [arcsec]      & [arcsec]     &[arcsec]&        &    \\
(1)      & (2)     & (3)     & (4)           & (5)          & (6)    &  (7)   & (8)      \\
\hline
 \noalign{\smallskip}

NGC 2983 & B$_a$      &  41.0    & 19.3          & 19.2       & 21.3  &0.0   & 0.543   \\ %.233
% \noalign{\smallskip} 
NGC 3081 & nb     & 117.6    &  5.7          &  5.6       &  7.8  &0.0   & 0.511   \\ 
         & AB     &  75.5    & 30.7          & 34.3       & 40.0  &1.8   & 0.382   \\ %barpa.288  
% \noalign{\smallskip} 
NGC 3100 & $\underline{\rm A}$B     & 163.8    & 27.0          & 31.7       &        &2.3   & 0.651 \\  % .288  
% \noalign{\smallskip} 
NGC 3166 & nb     &  86.5    &  2.3          & 3.1        &  6.8  &1.1   & 0.560 \\
\hline
\end{tabular}
\end{center}
\end{table}

\begin{table}
\begin{center}
%\caption{Continue  }
\begin{tabular}{llllllll}
\hline
 \noalign{\smallskip}         
Galaxy  & bartype & PA       &r$_{\rm vis}$ & r$_{\rm ell}$ & r$_{\rm L}$    & STdev  & q \\% ell=1-b/a -> b/a=1-ell     
        &         &[degrees] & [arcsec]& [arcsec] & [arcsec] &        &    \\
(1)     & (2)     & (3)      & (4)       & (5)      & (6)      &  (7)     & (8)      \\
\hline
 \noalign{\smallskip}

NGC 4691 & B      &  89.3    & 17.9          & 14.0       & 21.0  &1.9   & 0.289  \\ % .252
% \noalign{\smallskip} 
NGC 4694 & nb     & 145.2    &  3.2          &  2.5       &  3.4  &0.9   & 0.455  \\  % .25 
% \noalign{\smallskip} 
NGC 4754 & nb     &  18.9    &  8.1          &  6.7       &  9.6  & 0.7  & 0.754   \\ % .25 
         & B$_a$      & 131.5*   & 25.0          & 22.7       & 25.1  & 1.2  & 0.765 \\ 
% \noalign{\smallskip} 
NGC 4880 & $\underline{\rm A}$B     & 168.6    &  8.4          &  8.1       & 16.6  & 0.1  & 0.571  \\
\hline

\end{tabular}
\end{center}
*Visually estimated.
\end{table}

\clearpage
\newpage

%FIG 1 Example of flux calibrations

\clearpage
\newpage

%FIG 2 Flux cal coeff for diff campaigns

\clearpage
\newpage

%FIG 3 Flux calibration coeffients

\clearpage
\newpage

%FIG 4: image depth

\clearpage
\newpage

%\includegraphics[width=84mm]{FIG_1a.eps}
%\includegraphics[width=0.5\textwidth]{kcali_plot_good_bad.pdf}
% Add the following galaxies: 
%2782
%3384
%4245
%4369
%4382
%4472
%4546
%4552
%4638
%4649
%5353
%5493
%5846
%5898
%7029
%7585
%FIG 5; Atlas example:a) NGC2880
\begin{figure}
\includegraphics[width=170mm]{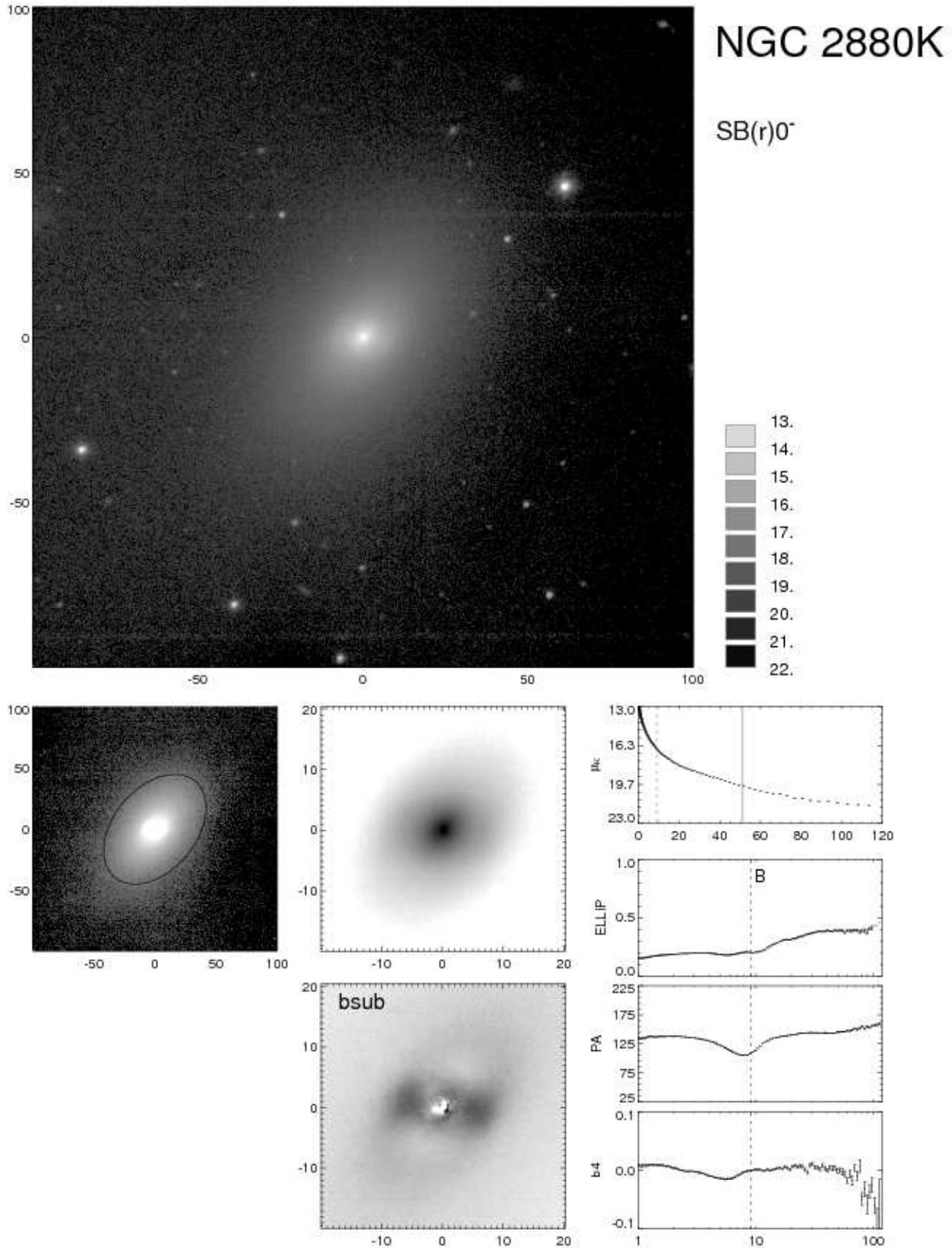}
\caption{An example of the atlas images, explained in more detail in the text
(for all galaxies: http://www.oulu.fi/astronomy/NIRS0S$\_$pub/atlas.html}
%\caption{NIRS0S atlas. {\it Upper papel:} the flux-calibrated images
% in the sky plane are given in units of mag arcsec$^{-2}$. The x- and
% y-scales in all frames are in arcseconds. {\it Lower image papels:}
% the images are shown in a linear scale, with the fore-grounds stars
% removed. In some cases residual images or unsharp masks (see Section
% XX) are also presented. {\it Right lower papels:} radial profiles of
% the surface brightnesses, ellipticities (ELLIP), position angles (PA)
% and of the deviations from perfect ellipticities ($a_4$) are
% shown. The dashed vertical lines mark the radii of the structure
% components assigned to our visual classification given in Table
% 2. The green vertical lines shows the radius where the surface
% brightness level is 20 mag arcsec$^{-2}$. The thick tick marks in
% y-axis show the orientation parameters used while deprojecting the
% images to face-on orientation.  The morphological types in the panels
% are our visual classifications in the near-IR. If the photometric
% type is different form the visual type, it is also given. }
\label{figure-5_1}
\end{figure}

\clearpage
\newpage

%NGC2782
\begin{figure}
\includegraphics[width=170mm]{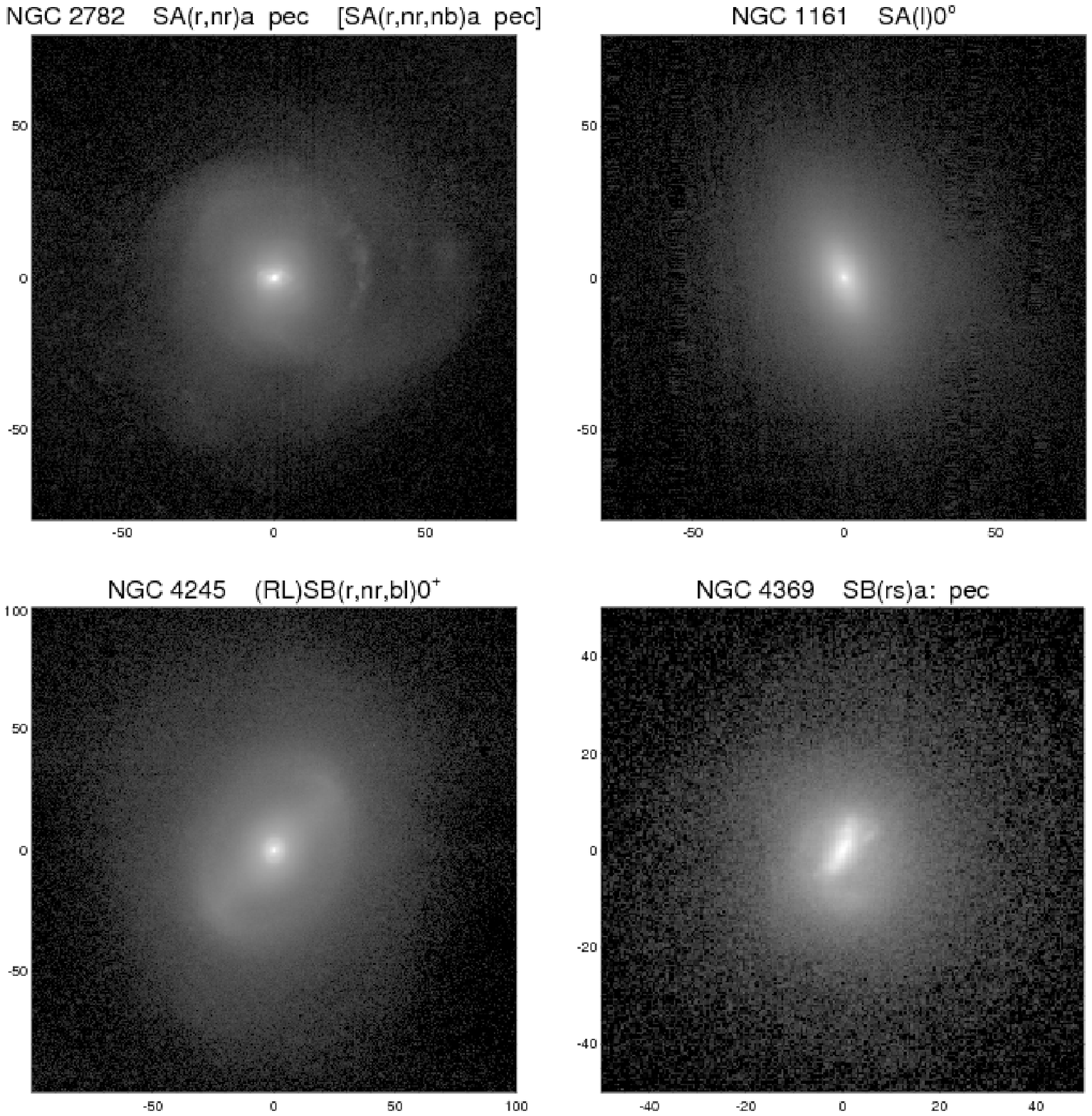}
%\caption{}
\label{figure-5_2}
\end{figure}

\clearpage
\newpage

%NGC3384
\begin{figure}
\includegraphics[width=170mm]{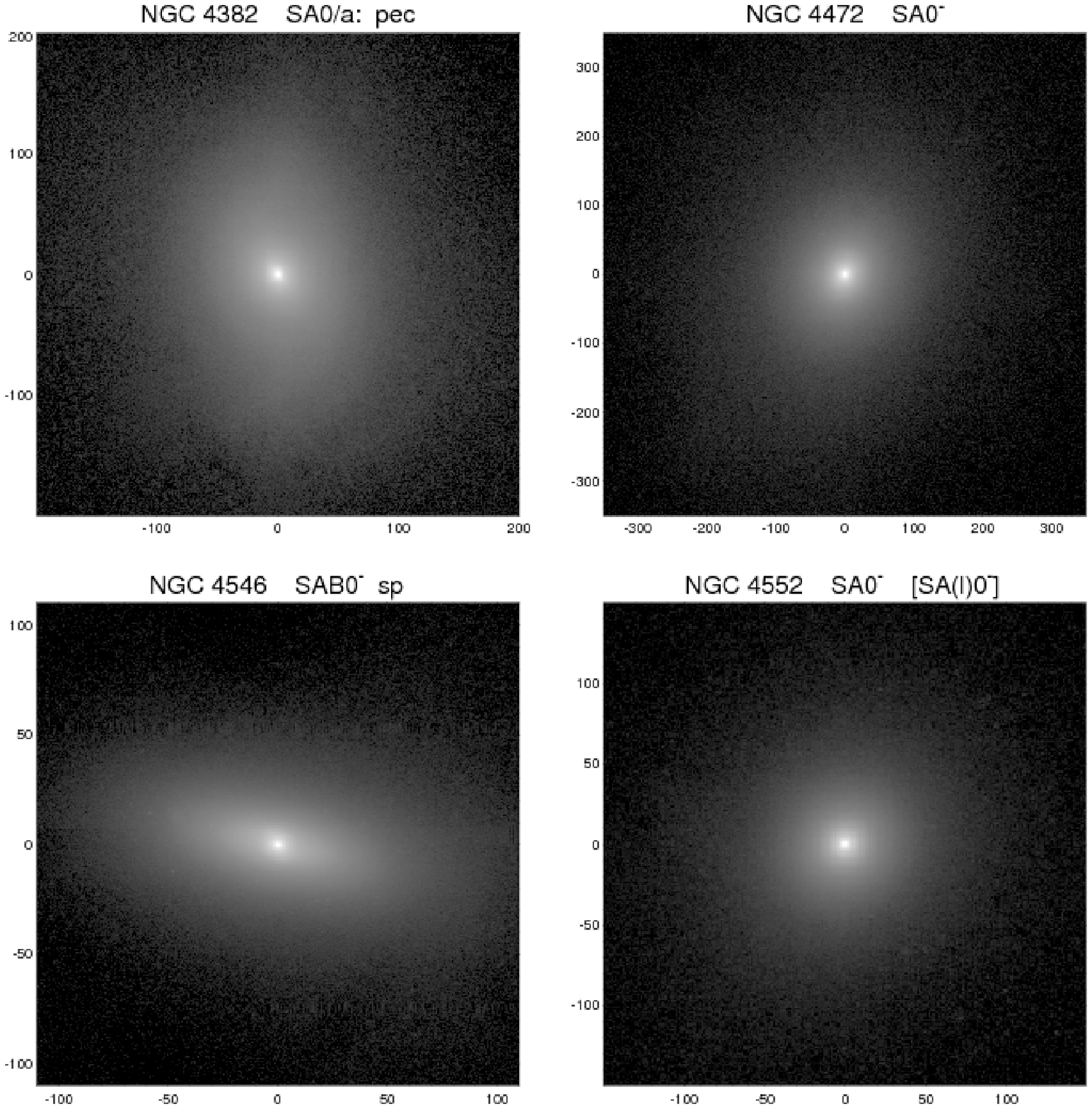}
%caption{}
\label{figure-5_3}
\end{figure}

\clearpage
\newpage
%NGC4245
\begin{figure}
\includegraphics[width=170mm]{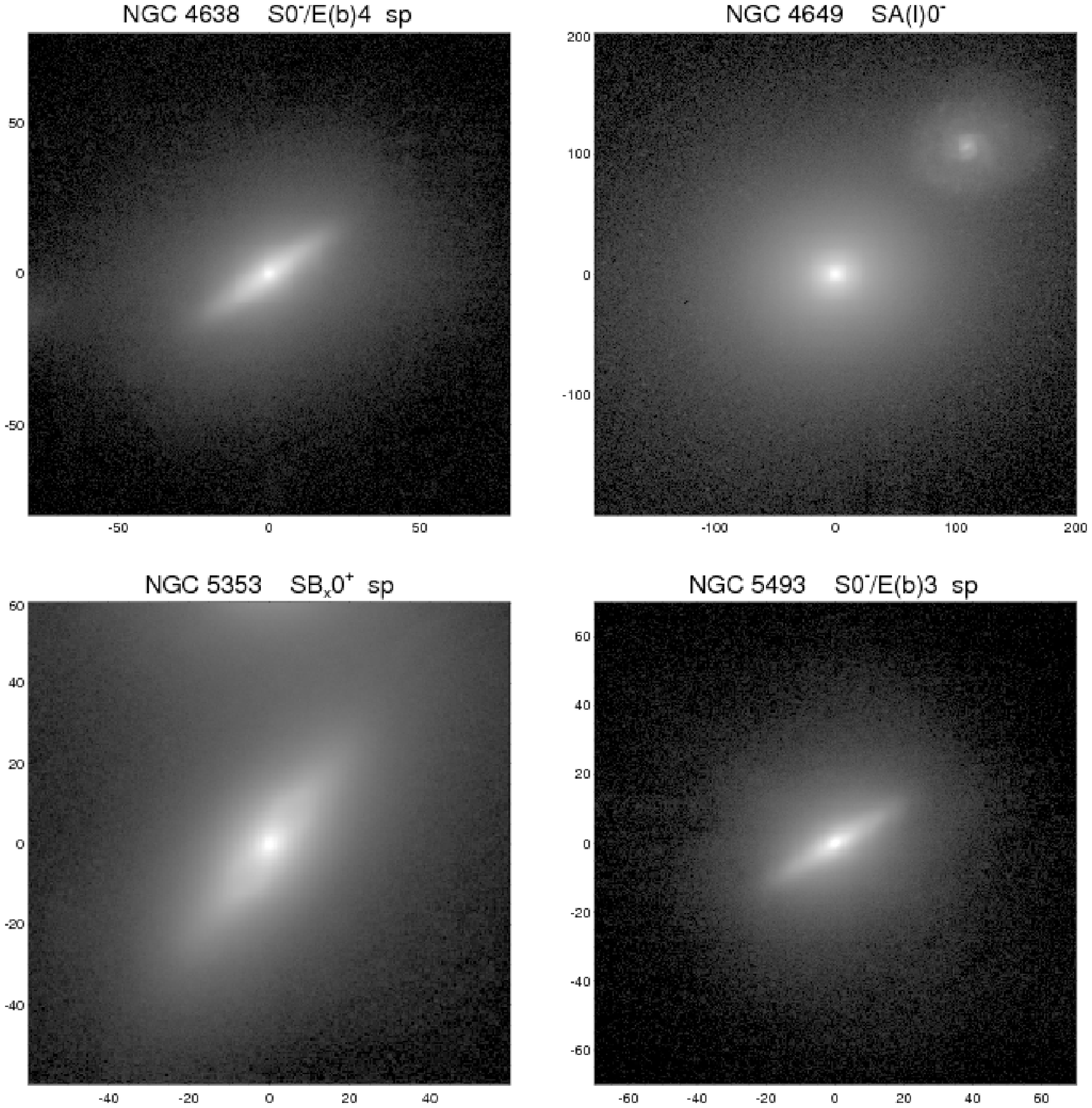}
%\caption{}
\label{figure-5_4}
\end{figure}

\clearpage
\newpage
%%NGC4245
\begin{figure}
\includegraphics[width=170mm]{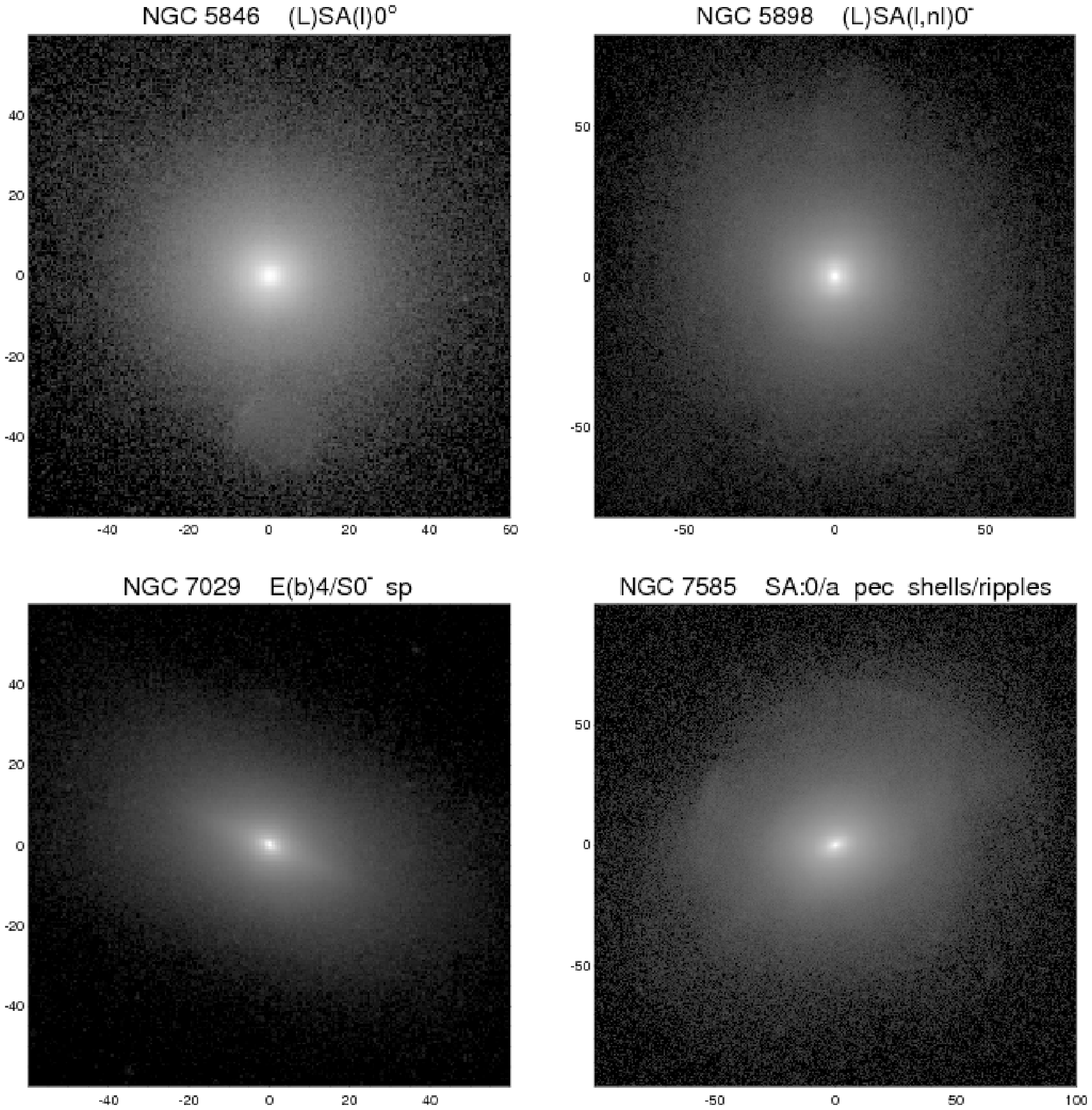}
%\caption{}
\label{figure-5_5}
\end{figure}

\clearpage
\newpage

%FIG 6: examples of stages
\begin{figure}
\includegraphics[width=150mm]{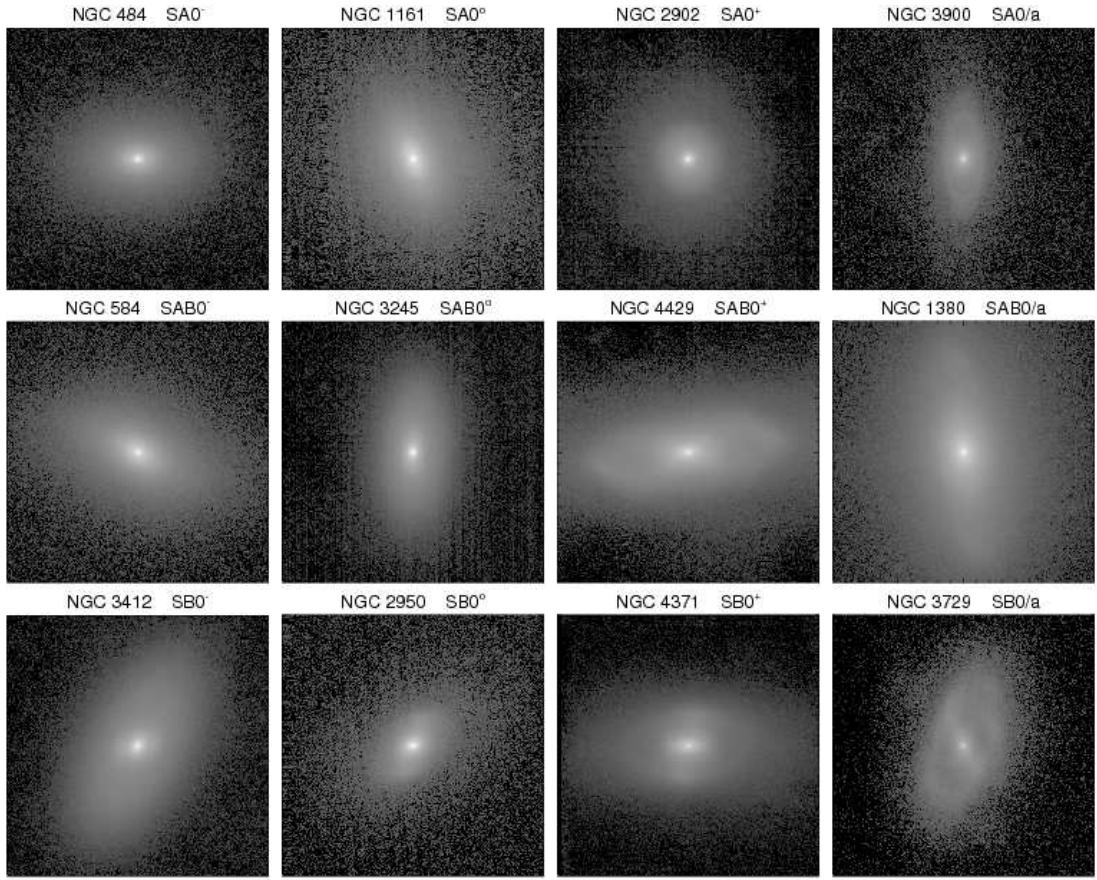}
\caption{Examples of {\it stage} (S0$^-$, S0$^o$, S0$^+$)
and {\it family} (A, SA, B) in the classification are shown. In this
and in the following figures the images are sky subtracted and they
are shown in a magnitude scale.}
\label{figure-6}
\end{figure}

%\clearpage
%\newpage

%FIG 7: bars and rings
\begin{figure}
\includegraphics[width=150mm]{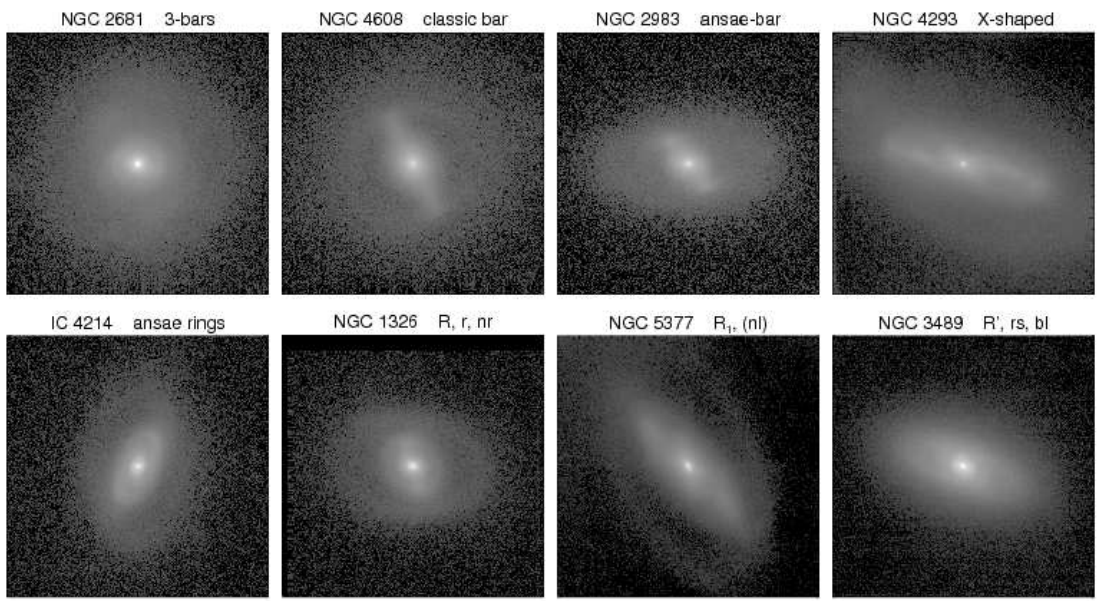}
\caption{Examples of bars and rings. The left upper panel (NGC 2681) shows
an example of a galaxy having three bars. The largest bar is
manifested as two weak ansae in the direction of 45 degrees
counter-clockwise from the North, whereas the main bar appears nearly
horizontally. The nuclear bar is visible only in the atlas image shown in Fig. 5. }
\label{figure-7}
\end{figure}

\clearpage
\newpage

%FIG 8: barlens in S0
\begin{figure}
\includegraphics[width=150mm]{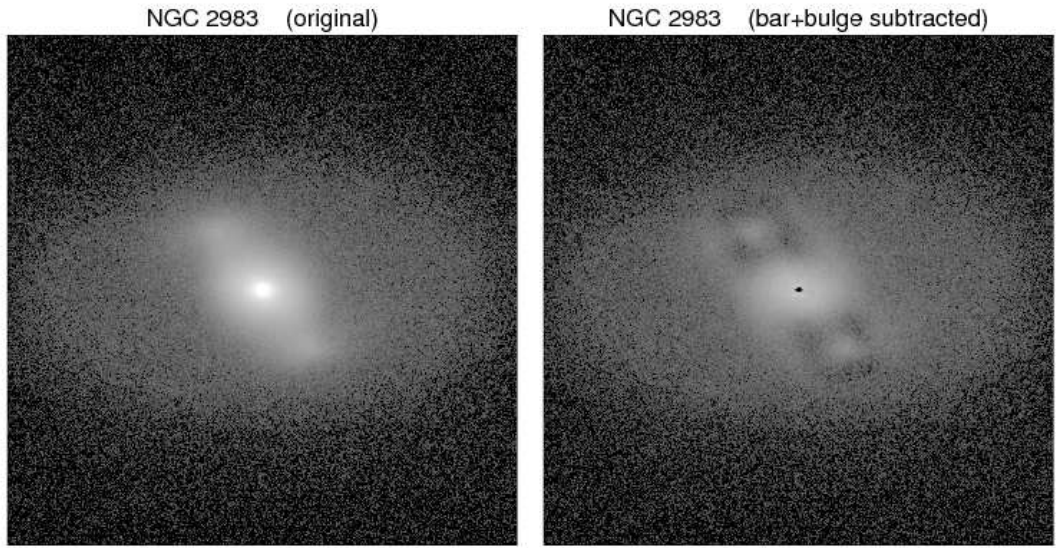}
\caption{ An illustration of barlens (bl) in NGC 2983. The left
panel shows the original image, and the right panel the residual image
after subtracting bar+bulge decomposition model taken from Laurikainen
et al. 2010). Barlens is the nearly spherical lens inside the bar. The
two blops (=ansae) at the two ends of the bar are real, but the ring-like
structure surrounding the bar is an artifact due to the fact that the
bar model is only an approximation of the true surface brightness
distribution of the bar.}
\label{figure-8}
\end{figure}

\clearpage
\newpage

\begin{figure}
\includegraphics[width=120mm]{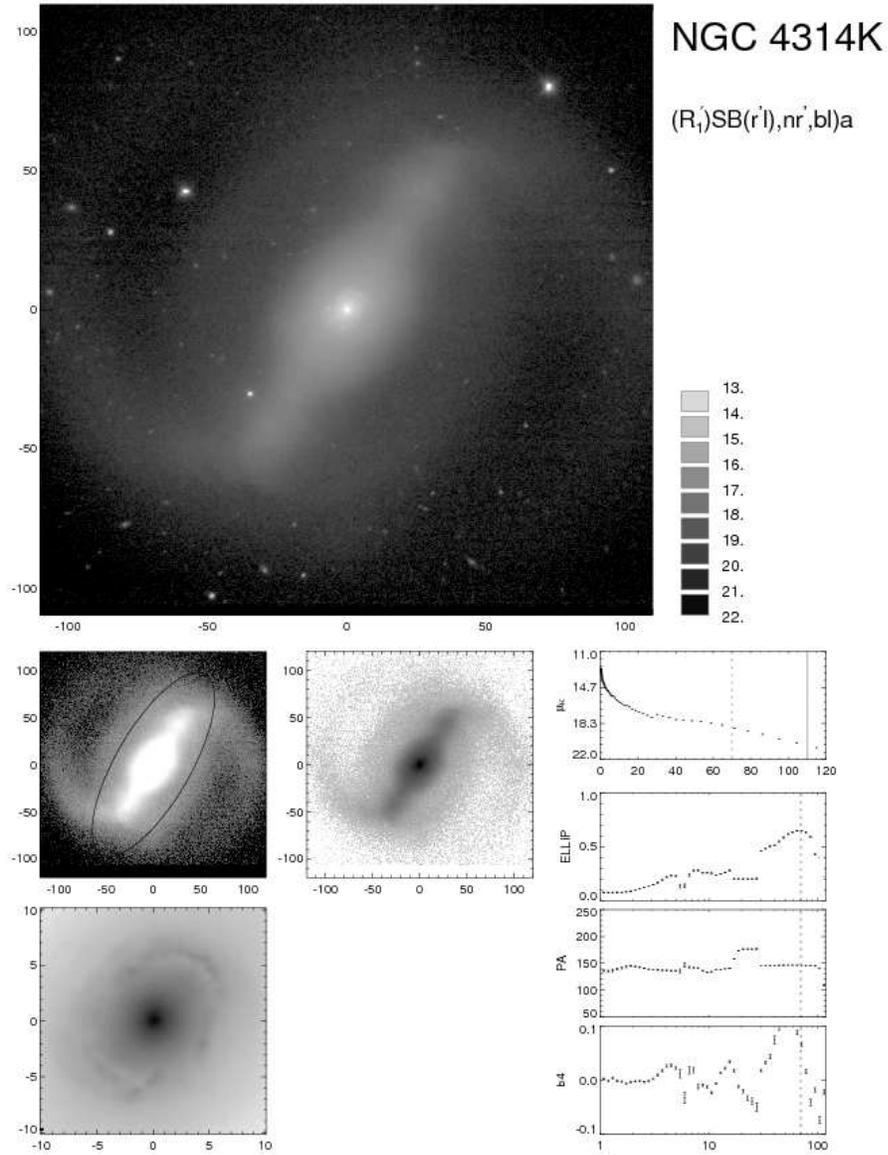}
\caption{ An illustration of barlens (bl) in an Sa-type spiral, NGC 4314: barlens is the
large component inside the bar, having a nuclear ring inside the barlens. The small panels 
are the same as in Figure 5.}
\label{figure-9}
\end{figure}

\clearpage
\newpage

%FIG_10
\begin{figure}
\includegraphics[width=120mm]{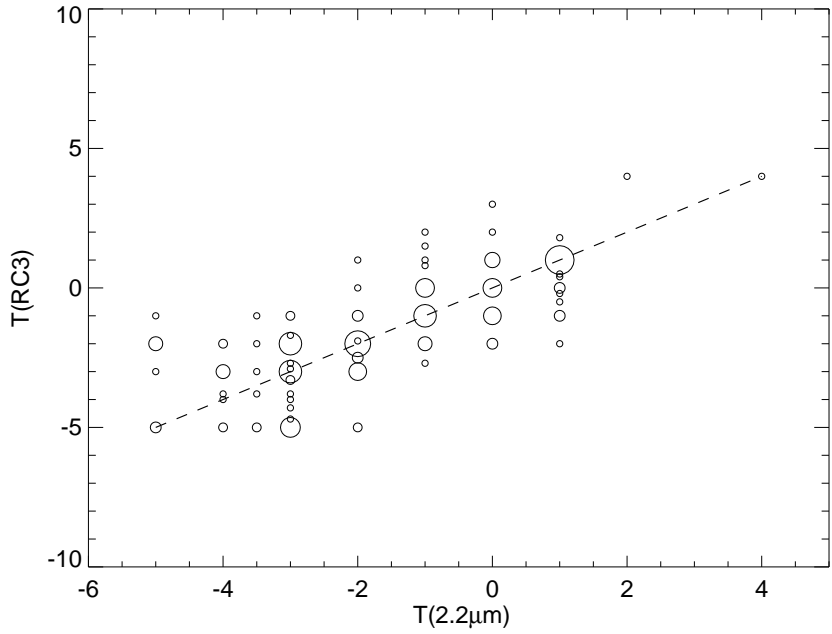}
\caption{The scatter plot of the optical and the NIR classifications, taken
from RC3 and this study, respectively. The size of the symbol indicates
the number of galaxies represented by the symbol. The dashed line corresponds to T(RC3)=T(2.2 $m \mu$)}
\label{figure-10}
\end{figure}

\clearpage
\newpage

%FIG 11: lenses
\begin{figure}
\includegraphics[width=110mm]{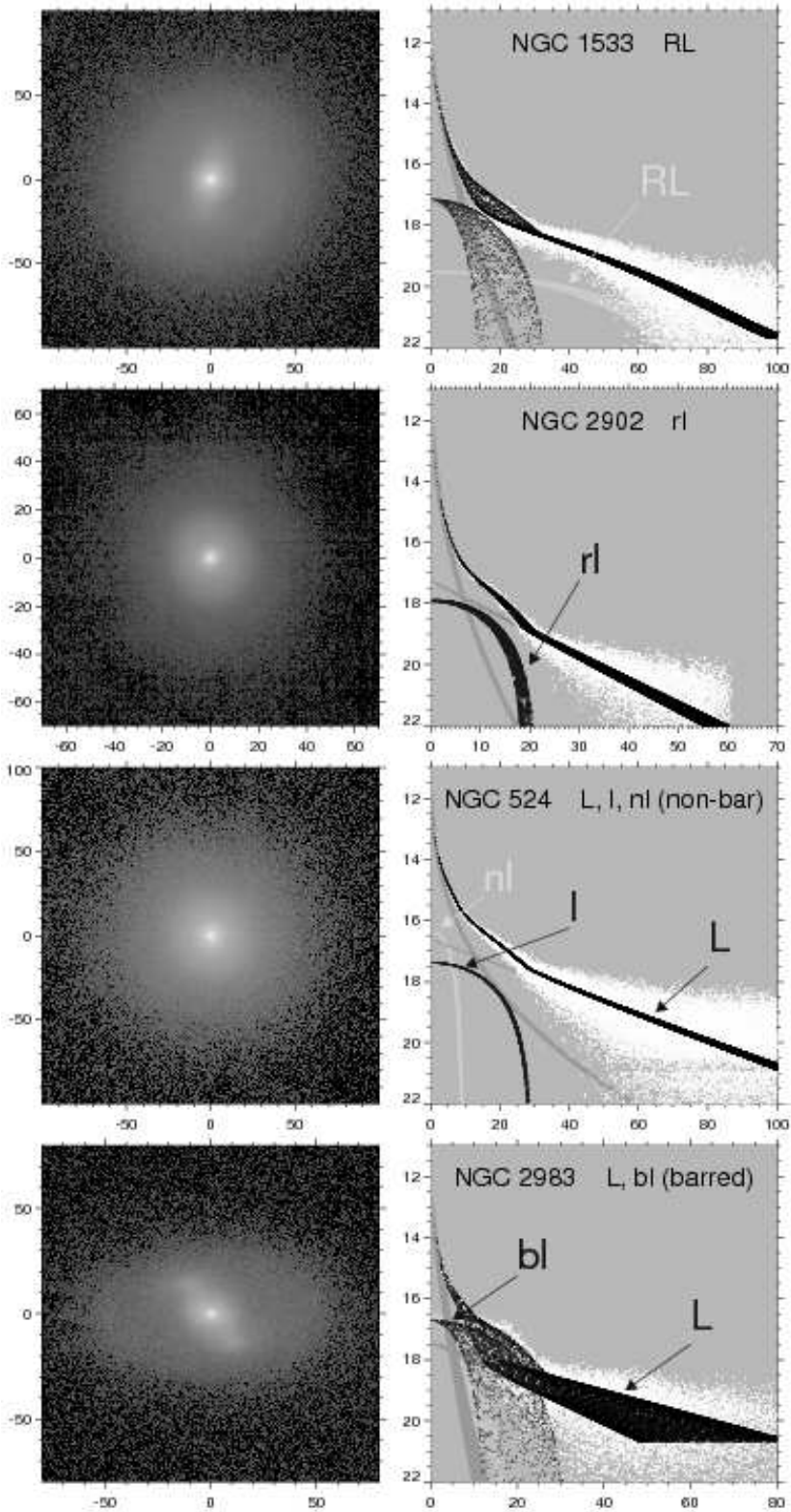}
\caption {Examples of different type of lenses. Left panels show
the images and right panels the 2D multi-component decompositions from
Laurikainen et al.  (2010), explained in more detail in the text. 
In the decomposition plots the white dots show the data points
  of the 2-dimensional surface brightness distribution (brightness of
each pixel as a function of sky-plane radius from the galaxy center), and the black
  and grey colors show the model components. The uppermost black dots show the total model.
%RL
%and rl are features which, when added on top of an exponential disk,
%makes the surface brightness profile curved. NGC 524 is a galaxy which
%has all 3 lens types, manifested as exponential subsections in the
%surface brightness profile. {}
}
\label{figure-11}
\end{figure}

\clearpage
\newpage

%FIG 12: photom class
\begin{figure}
\includegraphics[width=150mm]{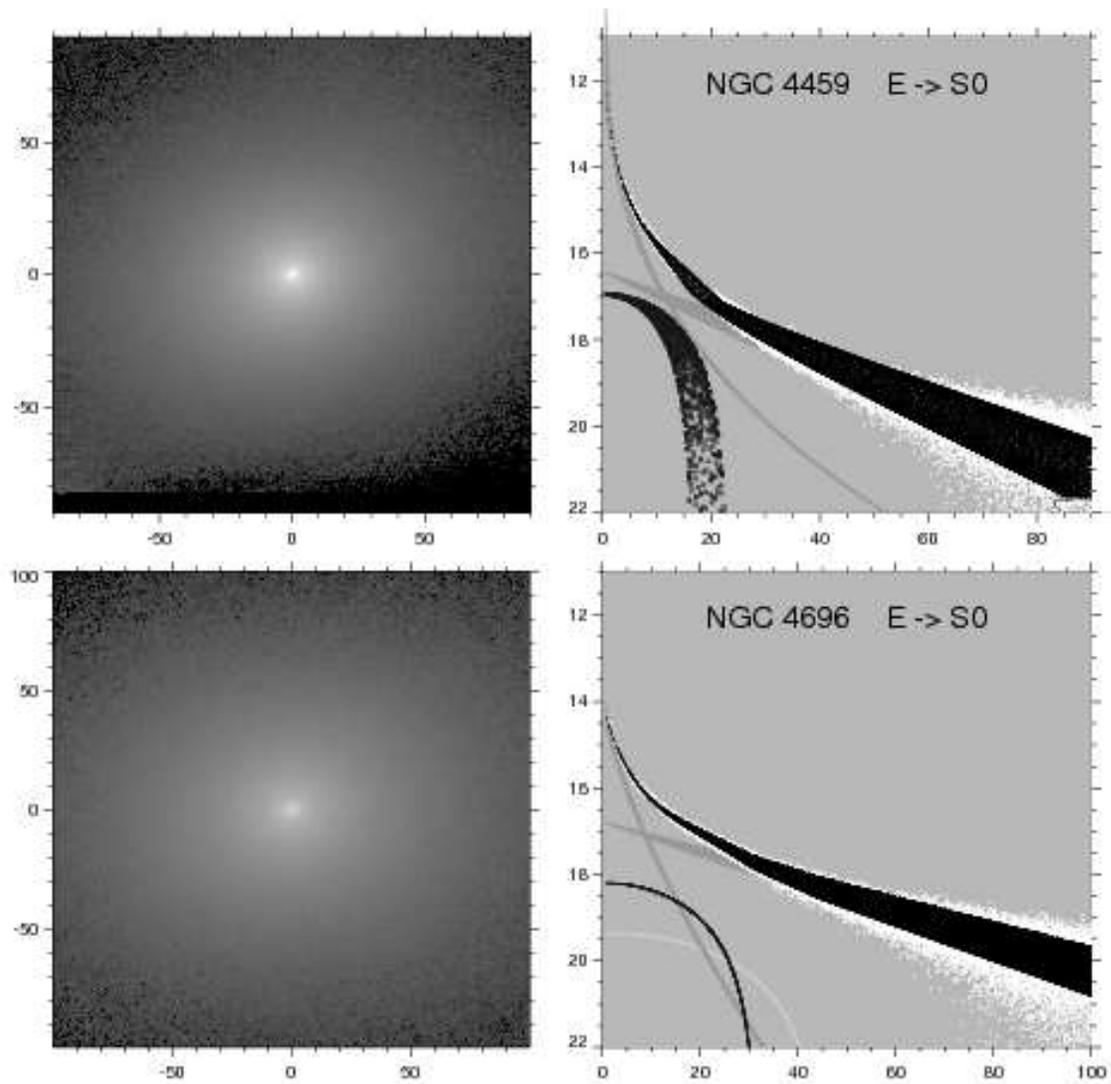}
\caption{Two examples in which photometric classification
moves an elliptical galaxy to an S0 stage. Both galaxies have an
exponential outer surface brightness profiles, and also evidence of 
lenses, manifested as exponential sub-sections in the brightness profile.
In the decomposition plots the meaning of the dots and lines are the same 
as in Fig. 10.}
\label{figure-12}
\end{figure}

\clearpage
\newpage

%FIG 13: measuring stragery of rings and lenses 
\begin{figure}
\includegraphics[width=100mm]{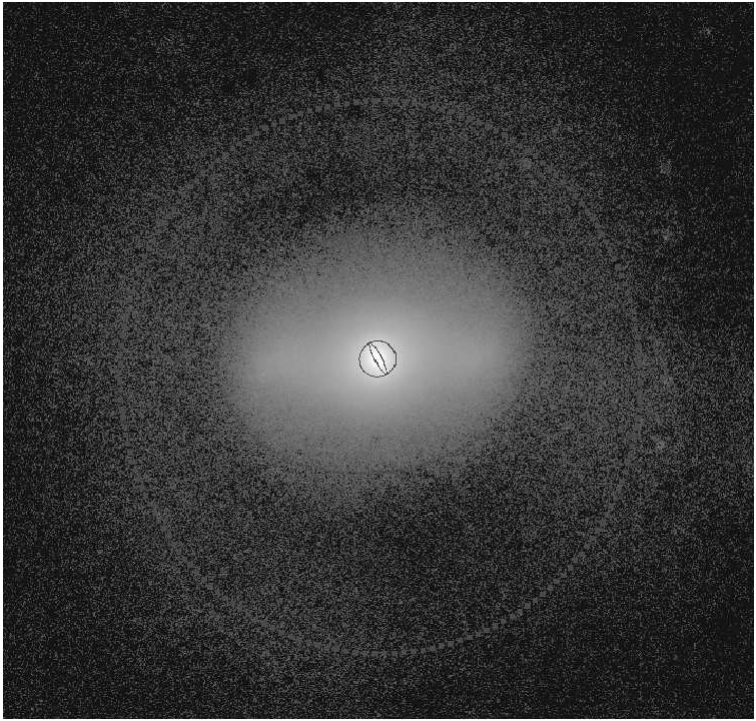}
\caption{ An illustration of our strategy for measuring the dimensions of the structures:
the fitted ellipsoids of the identified structures are shown for NGC 1543. From outside
towards inside the ellipsoids are for: the outer ring (R), the lens and the bar (l,B), and the 
nuclear lens (nl). Also the nuclear bar (nb) is fitted, though it has the same dimension
as the nuclear lens (for all galaxies: http://www.oulu.fi/astronomy/NIRS0S$\_$pub/nirsos$\_$dimensions.html).}
\label{figure-13}
\end{figure}

\clearpage
\newpage

%FIG 14: formative seq of lenses
\begin{figure}
\includegraphics[width=150mm]{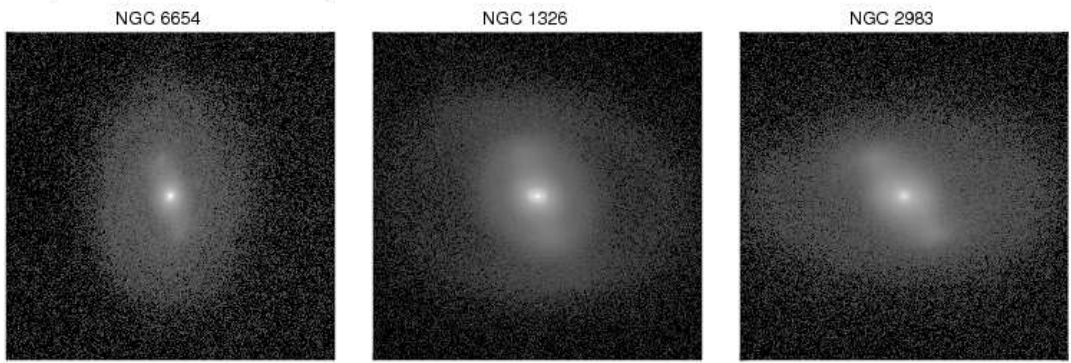}
\caption{Examples of possible formative sequences of lenses. In Fig. (a)
the two lower panels show the inner parts of the galaxies NGC 3998 and NGC 4203: in these
figures the bulge models obtained from the decompositions are subtracted
from the original images. }
\label{figure-14}
\end{figure}

\clearpage
\newpage

%FIG 15: multiple bars and lenses
\begin{figure}
\includegraphics[width=150mm]{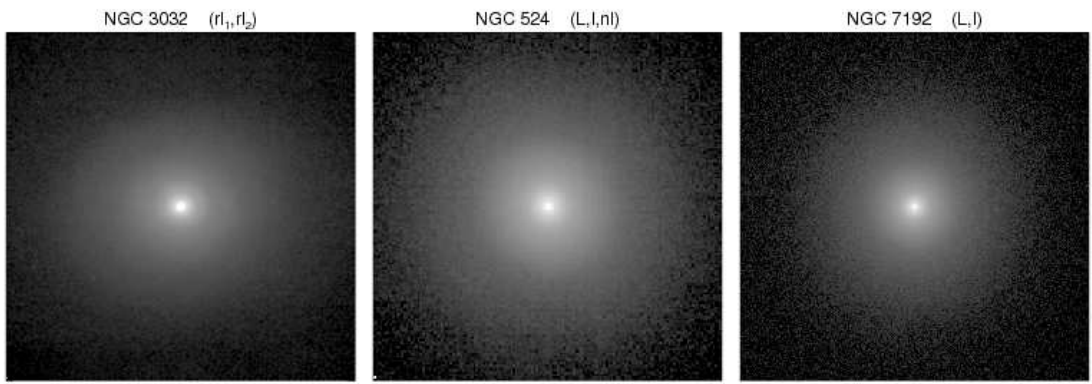}
\caption{ Examples of multiple lenses.
(a) Three barred galaxies are shown: the upper row illustrates the lenses surrounding
the primary bars, and the lower panels those surrounding the nuclear bars (only the
central regions of the galaxies are shown). For NGC 1543 and NGC 6782 the lenses are clear.
However, for NGC 3081 the nuclear and inner ring are very prominent and the two bars
extremely weak, so that no lenses are coded to the classification. (b) Typical 
examples of multiple lenses in non-barred galaxies. }
\label{figure-15}
\end{figure}

\clearpage
\newpage

%FIG 16: S0_c galaxies
\begin{figure}
\includegraphics[width=140mm]{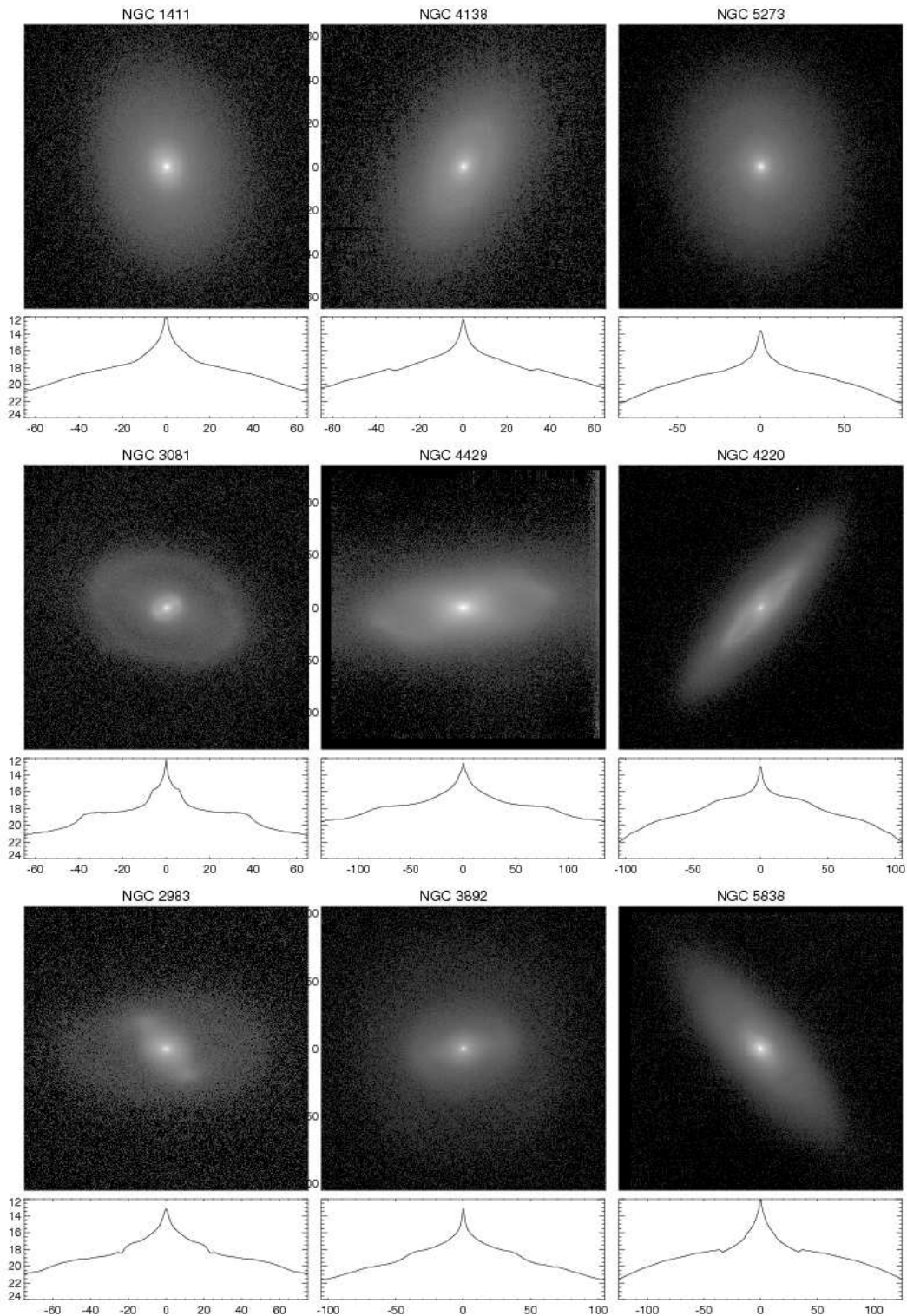}
\caption{Candidates of S0$_c$ type galaxies, showing no spiral arms
and bulge-to-total flux ratios as small as typically found for Sc-type spirals.
For each galaxy are shown the flux-calibrated cleaned image, and the azimuthally
averaged surface brightness profile. The radial scale is in arcsec.
 }
\label{figure-16}
\end{figure}

\end{document}